\documentclass[12pt]{article}
\usepackage{mathtools}
\usepackage{amsthm}
\usepackage{amssymb}
\usepackage{subcaption}
\usepackage{comment}
\usepackage{url}
\usepackage{multicol}
\usepackage{tikz}
\usepackage[a4paper, margin=1in]{geometry} 

\newtheorem{thm}{Theorem}

\begin{document}
\title{On the design and stability of cancer adaptive therapy cycles: deterministic and stochastic models.}

\author{Yuri G. Vilela$^{1}$\\ Artur C. Fassoni$^{2}$\\Armando G. M. Neves$^{1}$
	\\
	\normalsize{$^{1}$Department of Mathematics, Universidade Federal de Minas Gerais}
	\\ 	\normalsize{ygvilela@ufmg.br, aneves@mat.ufmg.br}\\
	\normalsize{$^{2}$Institute of Mathematics and Computing, Universidade Federal de Itajubá}	
	\\	\normalsize{fassoni@unifei.edu.br}
}

\date{\today}
\maketitle

  \begin{abstract}
        Adaptive therapy is a promising paradigm for treating cancers, that exploits competitive interactions between drug-sensitive and drug-resistant cells, thereby avoiding or delaying treatment failure due to evolution of drug resistance within the tumor. Previous studies have shown the mathematical possibility of building cyclic schemes of drug administration which restore tumor composition to its exact initial value in deterministic models. However, algorithms for cycle design, the conditions on which such algorithms are certain to work, as well as conditions for cycle stability remain elusive. Here, we state biologically motivated hypotheses that guarantee existence of such cycles in two deterministic classes of mathematical models already considered in the literature: Lotka-Volterra and adjusted replicator dynamics. We stress that not only existence of cyclic schemes, but also stability of such cycles is a relevant feature for applications in real clinical scenarios. We also analyze stochastic versions of the above deterministic models, a necessary step if we want to take into account that real tumors are composed by a finite population of cells subject to randomness, a relevant feature in the context of low tumor burden. We argue that the stability of the deterministic cycles is also relevant for the stochastic version of the models. In fact, Dua, Ma and Newton [Cancers (2021)] and Park and Newton [Phys. Rev. E (2023)] observed breakdown of deterministic cycles in a stochastic model (Moran process) for a tumor. Our findings indicate that the breakdown phenomenon is not due to stochasticity itself, but to the deterministic instability inherent in the cycles of the referenced papers. We then illustrate how stable deterministic cycles avoid for very large times the breakdown of cyclic treatments in stochastic tumor models.
    \end{abstract}
    
   \textbf{Keywords:} Cancer Therapy; Mathematical Oncology; Evolution of Resistance; Lotka-Volterra; Replicator dynamics; Moran process

	\section{Introduction}
	
	Despite continuous research efforts and significant advancements in its treatment over the years, cancer remains one of the major causes of premature death worldwide \cite{CancerStats}. Drug resistance, in particular, has proven to be a formidable challenge in this field, affecting every known drug and being responsible for over $90\%$ of cancer-related deaths among chemotherapy patients \cite{Aktipis, Bukowski}. 
 
\begin{figure}[ht]
		\centering
		\includegraphics[width=0.8\textwidth]{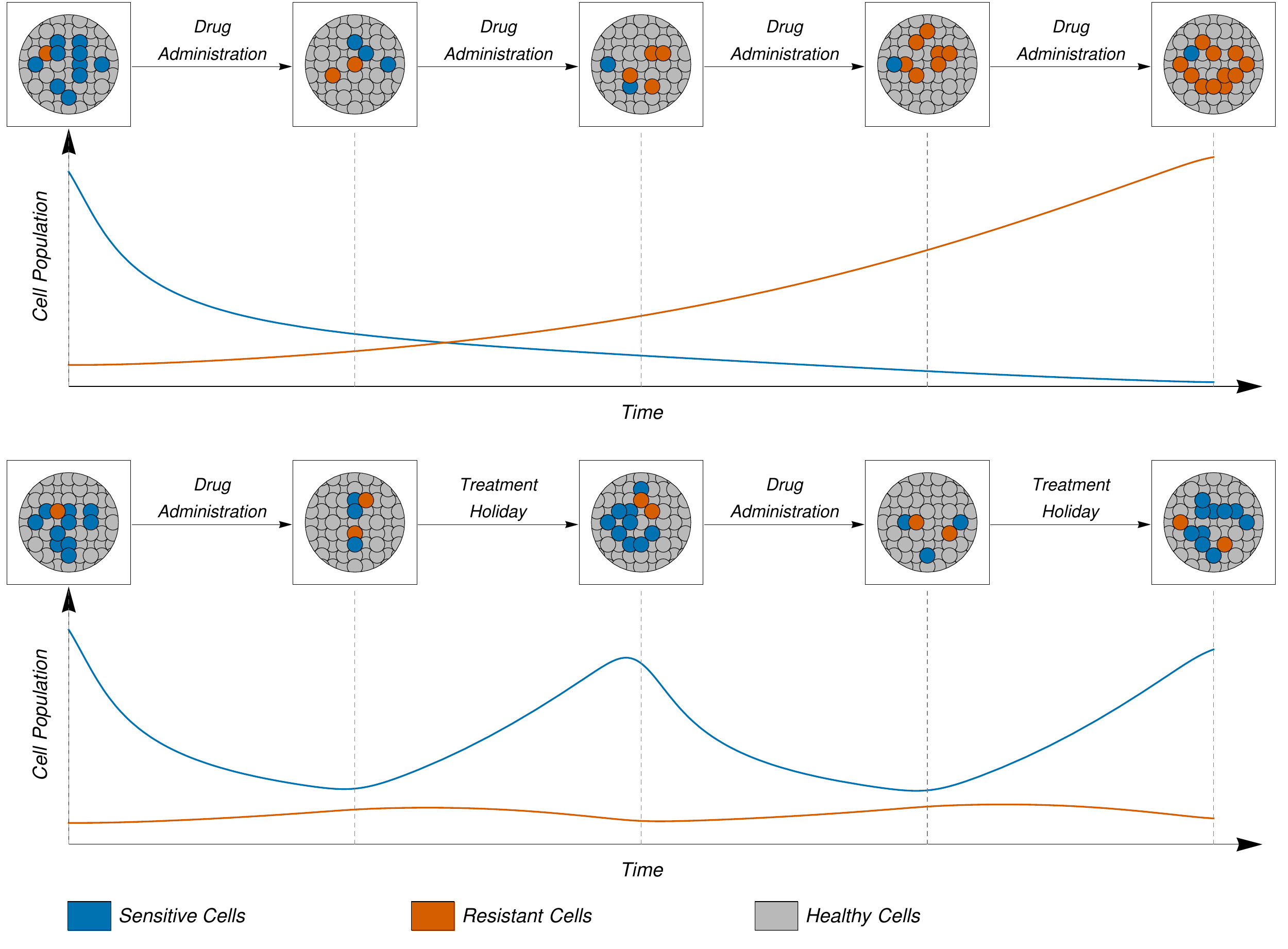}

		\caption{Representation of a continuous treatment (above) and a treatment routine including treatment holidays (below). While continuous treatment leads to the evolution of resistant cells due to competitive release, intermittent/adaptive therapy with drug holidays allows the competition with sensitive cells to suppress the growth of resistant cells, aiming to keep the tumor under control instead of eradicating it.}
        \label{fig:continuousVsAdapt}
	\end{figure}

    An evolutionary mechanism based on the concept of competitive release explains the evolution of resistance \cite{Aktipis}. Due to the highly heterogeneous environment, resistant cancer cells are often already present in the tumor prior to any treatment, but are naturally suppressed by a fitter population of sensitive cells in the absence of the drugs. However, as treatment proceeds, strong selective pressure is applied to the tumor, creating an environment favoring the rise of a resistant population. Therefore, if the treatment does not weaken the tumor enough so that non-cancerous cells are able to control it by themselves, a new, predominantly resistant, tumor emerges. This is illustrated in Fig. \ref{fig:continuousVsAdapt}.
	
	Borrowing concepts from pest control in agriculture, \textit{adaptive therapy} \cite{Gatenby} proposes a ``control over elimination" approach, exploiting the cost of resistance and the naturally competitive environment of the tumor to manage its growth, allowing patients to coexist with the disease for extended time periods. To achieve this, personalized treatment routines are designed to include ``treatment holidays," i.e. periods without any drugs, in order to allow the sensitive cell population to recover and suppress the emergence of resistant cells. An illustration for that can be seen in the bottom row of Fig. \ref{fig:continuousVsAdapt}. Treatments with adaptive therapy have already begun to be translated into preclinical trials for various cancer types, showing promising results and retarding tumor progression \cite{Adler, Gedye, Zhang2023}.
	
	From a mathematical standpoint, several \textit{deterministic models} have been proposed over the years, encompassing both single-drug \cite{Zhang2017, West2018Replicator, Wang2021} and multidrug \cite{West2019LotkaVolterra, West2020, DuaMaNewton2021} adaptive therapy schemes. While the simplicity of single-drug models greatly facilitates their analysis, most real-life treatments involve multiple drugs, making multidrug models more representative of reality. Furthermore, the use of multiple therapies allows for greater control over the tumor by targeting different cell populations at different times, facilitating the creation of \textit{cyclic treatment routines}. These routines intend to restore roughly the same tumor composition before and after their application, making them treatments that can, at least in theory, be conducted indefinitely \cite{West2020}.	
	
	From a deterministic perspective, the cell populations within a tumor are mathematically modeled by a system of ordinary differential equations (ODEs) containing several parameters, and the effects of different drugs are dealt with by different values of the parameters. The construction of a cyclic routine amounts to finding \textit{exact} durations for the treatment with each drug, so that the the solution of the ODE system returns to the \textit{exact} point where it was at the beginning of the cycle. Although the construction of such cycles may seem straightforward, the underlying mathematical assumptions for its success must be made explicit. Specifically, when multiple cell types are accounted for in the model, the design of a cyclic routine becomes a matter of intersecting curves (orbits of an ODE system) in a plane, or in higher-dimensional spaces. Even in planar contexts, the existence of such cycles is not guaranteed without mathematical assumptions. Throughout this paper we establish formal, yet biologically motivated, conditions that guarantee the existence of such cycles in two different deterministic models: Lotka-Volterra equations \cite{hofbauer1998evolutionary, murray} and adjusted replicator dynamics \cite{taylorjonker, hofbauer1998evolutionary, Nowak2006}.
	
	Moreover, even after successfully designing a cyclic routine, another critical consideration is its \textit{stability}, i.e., how closely the actual ODE systems solutions mirror the intended ones in case of small errors in the initial condition, due e.g. to the inherent modelling approximation and inevitable experimental errors in its determination. Arguably, a stable routine could be reliably employed over prolonged periods, while unstable ones would require frequent adjustments to be effective. 
	
	As shown in this paper, stability of deterministic routines plays an important role also when considering \textit{stochastic models} for tumors. Recent studies \cite{DuaMaNewton2021, ParkNewton2023}, motivated by the fact that population cells are always finite and subject to random fluctuations, particularly relevant within the context of low tumor burden, considered robustness of cyclic deterministic routines produced within a deterministic model (adjusted replicator dynamics) when this model is replaced by its \textit{stochastic counterpart}, the Moran process \cite{moran, Nowak2006}. They observed the collapse of deterministic cycles in the stochastic counterparts due to the stochastic fluctuations that quickly deviate the cycles.
	
	Of course, due to the very stochastic nature of the Moran process, one should not expect that a stochastic routine derived from a deterministic cycle will return exactly to its starting point. We can expect it to return, with large probability, \textit{close} to its starting point, if the population size $N$ is large enough. If the deterministic cycle is stable, we will see that when it is realized in the stochastic context, the tumor composition remains close to the intended deterministic solution for many cycles. On the contrary, \textit{if the cycle is unstable}, we expect that the stochastic tumor composition will break down after a few cycle repetitions. We will show that this breakdown, already observed in \cite{DuaMaNewton2021, ParkNewton2023}, is here correlated with the instability of the underlying deterministic cycle. Besides the Moran process, which is a stochastic counterpart of the adjusted replicator dynamics, we will also study a stochastic counterpart for the Lotka-Volterra dynamics.

    This paper is organized as follows. In Section \ref{sec:notations}, in order to make the paper more self-contained, we briefly introduce some mathematical terms and notations to be used throughout. In Section \ref{sec:deterministicModels}, we present two deterministic models for tumor dynamics: the Lotka-Volterra model (Subsection \ref{subsec:LVModel}) and the adjusted replicator model (Subsection \ref{subsec:repModel}). For each model, we provide biologically relevant conditions that guarantee the existence of  cyclic routines starting at a given point. We show that, in fact, infinitely many of these cycles exist and exhibit a simple method to design them. We introduce then (Subsection \ref{subsec:Stability}) the concept of stability and give examples of stable and unstable routines both for Lotka-Volterra and adjusted replicator dynamics. In Section \ref{sec:stochasticModels}, we present stochastic counterparts to both deterministic models and examine the roles of population size and cycle stability in the expected behavior when random fluctuations are considered. The paper is closed by concluding remarks in Section \ref{sec:conclusion} and two appendices containing more mathematical material.

	 \section{Some mathematical terminology and notations}
  \label{sec:notations}
    
Throughout this paper, the cell composition of a tumor with $n$ different cell types will be denoted by vectors $x = (x_1, \cdots, x_n)$ in the $n$-dimensional space $\mathbb{R}^n$. The $i$-th coordinate of $x$ is denoted by $x_i$ and represents, depending on the model being studied, either the number or the fraction of cells with type $i$ in the tumor.

In the deterministic context, cell populations within a tumor will be modeled by autonomous ODE systems $x^\prime = F(x)$. In all cases, for a given initial condition $x$ at time $t=0$, the ODEs appearing in this paper have unique solutions for all times both in the future and in the past. The \textit{flow} $\Psi(t, x)$ of an ODE system is the function that associates to each pair $(t, x)$ the solution of the ODE with initial condition $x$ at time $t$. The \textit{orbit} of a point $x$ is the curve in $\mathbb{R}^n$ of all points reached by the flow $\Psi(t,x)$ as $t$ is varied. Sometimes we will refer to the \textit{future} and \textit{past} orbits of a point as well, i.e. the curves defined by the flow with only values for $t \geq 0$ and $t \leq 0$ respectively. The \textit{flow diagram} of an ODE system is a picture of some orbits representative of the whole dynamical behaviour of the system.

\begin{figure}[ht]
    \centering
    \includegraphics[width=0.5\linewidth]{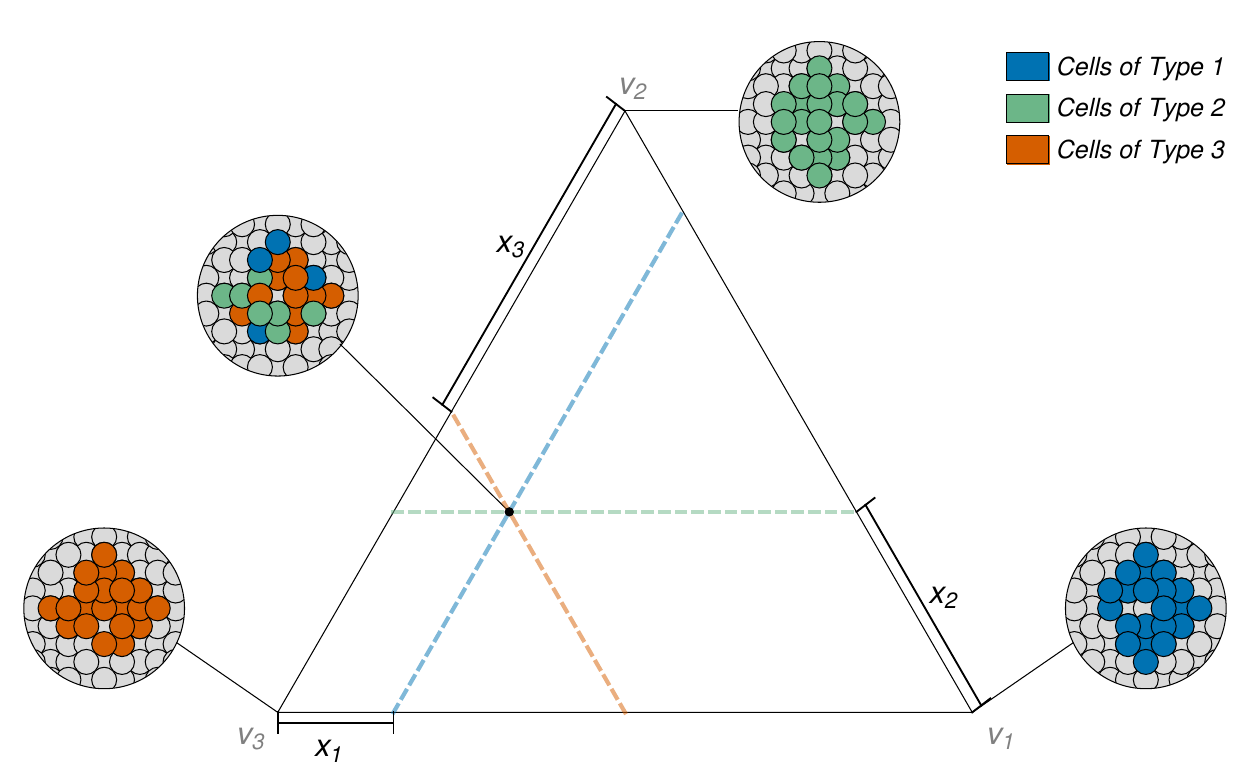}
    \caption{Simplex representation of different tumor composition possibilities.}
    \label{fig:simplexSample}
\end{figure}

We will consider in Subsection \ref{subsec:repModel} the adjusted replicator dynamics. Within this context, the $i$-th coordinate of a vector $x$ is the population fraction of cells with type $i$. Naturally, it is necessary that for $i=1, 2, \dots, n$ we have $0 \leq x_i \leq 1$ and the fractions must sum to 1, i.e. $x_1+ \dots + x_n=1$. Although $x$ is a vector in a $n$-dimensional space, the adjusted replicator dynamics is restricted to a $(n-1)$-dimensional part of it, namely the \textit{simplex}
$$S_{n-1} = \{x \in \mathbb{R}^n: x_1, x_2, \dots,  x_n \in [0, 1]; x_1 + \dots + x_n = 1\}\;.$$
In the particular case $n=3$ to be considered in this paper, $S_2$ is an equilateral triangle in three-dimensional space, but it will be represented as an equilateral triangle in the plane with sides equal to 1 length unit. The three vertices of the triangle are labelled by $V_1$, $V_2$ and $V_3$ and represent the population states in which all cells are of the same type as the vertex label. The sides of the triangle represent the states in which one of the cell types is absent. For example, the states in which only types $1$ and $2$ are present in the population are represented by points in the $V_1V_2$ side. More generally, the point $(x_1, x_2, x_3)$ of $S_2$ is represented by the point in the triangle at the intersection of lines parallel to the sides of the triangle, as shown in Fig. \ref{fig:simplexSample}. 

As usual in Mathematics, $\mathring{S_2}$ will be used to denote the \textit{interior} of $S_2$, i.e. the set of points in $S_2$, but not in its boundary.

In Subsection \ref{subsecMoran} we will consider the Moran stochastic process in a population with three types of cells. In the Moran process the total number of cells is a constant $N$. If $x_i$ denotes the number  of type $i$ cells, then $x_1+x_2+x_3=N$ and in figures concerning the Moran process we will represent the population state again by a point $(\frac{x_1}{N}, \frac{x_2}{N}, \frac{x_3}{N})$ in the triangle $S_2$. 

    \section{Deterministic Models}
    \label{sec:deterministicModels}
    Let us consider the deterministic dynamics of a tumor with $n$ cell types subject to $m$ different treatment phases. Each phase will be modelled by a different ODE system, so that we have $m$ ODE systems 
    $$x^\prime = F_1(x) \;, \dots \;,  x^\prime = F_m(x)$$
    with respective \textit{flows} $\Psi_i(t, x)$, $i=1, 2, \dots, m$.  We will denote by $T_1,\cdots,T_m$ the treatment phases durations. 

    A routine composed by the above treatments, starting with the tumor composition $x^*$, is \textit{cyclic} if the sequential application of the $m$ treatments with the corresponding durations returns to the original state, i.e., if $$(\Psi_m(T_m, \ \cdot\ ) \circ \cdots \circ \Psi_1(T_1, \ \cdot\ ))(x^*) = x^*.$$

    Let the initial point $x^*$ and durations $T_1, \cdots, T_{m-2}$ for the $m-2$ first treatment phases be given. Define 
    $$y^* = (\Psi_{m-2}(T_{m-2}, \ \cdot\ ) \circ \cdots \circ \Psi_1(T_1, \ \cdot\ ))(x^*)$$
    as the tumor state after the first $m-2$ treatment phases. A putative algorithm for constructing a cyclic routine consists of determining the durations $T_{m-1}$ and $T_m$ of the last two phases such that the future orbit of $y^*$ under the $(m-1)$-th treatment intersects the past orbit of $x^*$ under the $m$-th treatment, i.e.
    $$\Psi_{m-1}(T_{m-1},y^*)= \Psi_{M}(-T_{m},x^*)\;.$$ 
    Finding $T_{m-1}$ and $T_m$ can be done numerically using any ODE solver (e.g. 4th order Runge-Kutta \cite{Butcher}).

    Of course, in general these times and intersection may not exist, and in such cases it is impossible to build cyclic routines from the given point with the desired treatment durations. In the following subsections, we will present, in two planar ODE models with $m=3$ (i.e. three treatments), conditions that guarantee the existence of cyclic routines determined by the above algorithm. In fact, we will show that whenever a cyclic routine starting at a given point exists, an infinite number of such routines can be designed, which enables the possibility of using optimization techniques to choose a best routine according to some criterion.
 
	\subsection{Lotka-Volterra Model} \label{subsec:LVModel}
	
	We start our studies studying a tumor composed of two cell types, type 1 (\textit{sensitive} cells) and type 2 (\textit{resistant} cells) treated with two different drugs, $D_1$ and $D_2$. We assume that type 1 cells are affected by both drugs while type 2 cells are resistant to $D_1$ but sensitive to $D_2$, and consider routines composed of three phases:	
	\begin{itemize}
		\item \textbf{Phase 1} is a treatment holiday, so no drug is used;
		\item During \textbf{phase 2} drug $D_1$ is used;
		\item During \textbf{phase 3} drug $D_2$ is used.
	\end{itemize}
	
	Let $x_1$ and $x_2$ stand for the population sizes of cell types 1 and 2. We model this scenario using \textit{competitive Lotka-Volterra} \cite{hofbauer1998evolutionary, murray} equations. The dynamics in any phase $i = 1, 2, 3$ is described by the ODE systems \begin{equation}
		\begin{cases}
			x_1^\prime = \displaystyle r_{1, i}x_1\left(1 - \frac{x_1 + a_{12, i}x_2}{k_{1, i}}\right) \\
			x_2^\prime = \displaystyle r_{2, i}x_2\left(1 - \frac{a_{21, i}x_1 + x_2}{k_{2, i}}\right) \\
		\end{cases} \;,
		\label{eqs:systemLV}
	\end{equation} 
 where the parameters values depend on the phase: $r_{j, i} > 0$ and $k_{j, i} > 0$ are respectively the Malthusian growth rate and the carrying capacity of type $j$ cells in the $i$-th treatment phase, and $a_{jl, i} > 0$ is aggressiveness of type $l$ cells towards type $j$ cells at phase $i$.
	
\begin{figure}[ht]
		\centering
		
		\includegraphics[width=\linewidth]{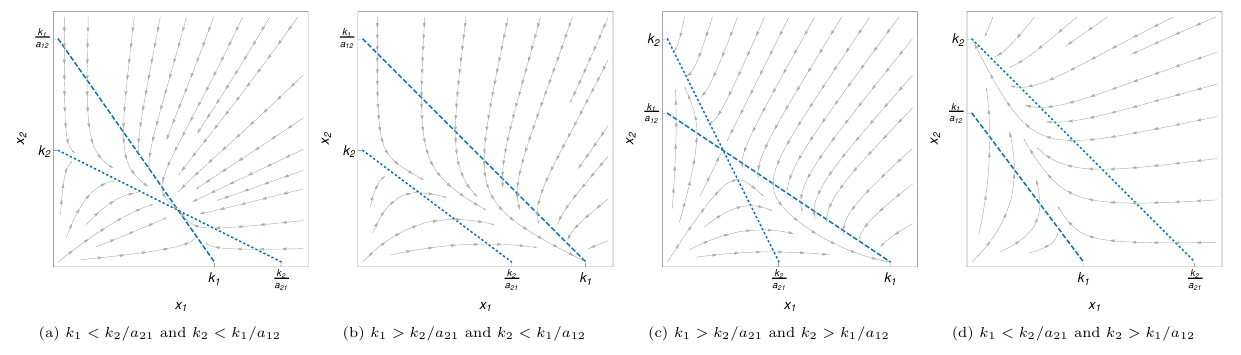}
		
		\caption{Flow diagrams of Lotka-Volterra systems with stable coexistence (a), sole survival of $x_1$ (b), unstable coexistence (c) and sole survival of $x_2$ (d). In all four cases, the dashed line is the $x_1$ nullcline and the dotted line is the $x_2$ nullcline.}
		\label{fig:possibleLV}
	\end{figure}

	Despite their generality, the dynamics of such systems can be easily classified by analyzing the relative position of their nullclines, i.e. the curves where the right-hand side of each equation is null. For a generic two-dimensional competitive Lotka-Volterra system, the nullclines are concurrent lines and there are four possible dynamical behaviors (see, for instance, section 3.3 of \cite{hofbauer1998evolutionary}): stable (Fig. \ref{fig:possibleLV}a) and unstable coexistence (Fig. \ref{fig:possibleLV}c) if the nullclines intersect in the first quadrant, or, otherwise, the sole survival of a single (\textit{dominant}) species regardless of the initial condition (Figs. \ref{fig:possibleLV}b and \ref{fig:possibleLV}d).
	
\begin{figure}[ht]
	
	\centering

    \includegraphics[width=\linewidth]{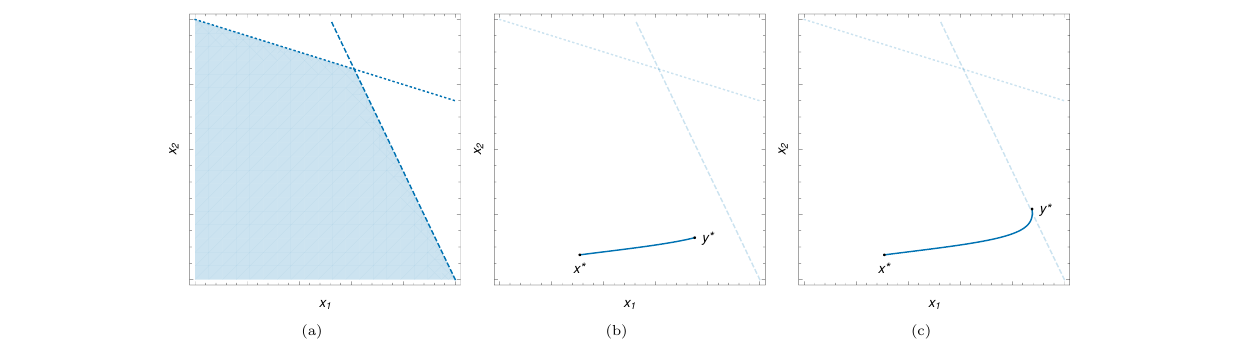}
	
	\caption{Region $R_1$ (a) and the orbit of $x^*$ during the first phase of the treatment (blue curve) with different duration times $T_1$ and their respective endpoints $y^{*} = \Psi_1(x^*, T_1)$ (b,c).}
    \label{fig:phase1Lv}
\end{figure}

In all four cases in Fig. \ref{fig:possibleLV}, populations of both cell types grow in the region below the two nullclines. We define $R_1$ as the  region  below the two nullclines in the first phase of the treatment, i.e. when no drugs are administered, see Fig. \ref{fig:phase1Lv}a. Points in $R_1$ can be interpreted as tumors where populations of both cell types are still increasing, and are natural candidates to start a treatment routine.

If we start our routine with $x^* \in R_1$, then the orbit $\Psi_1(t, x^*)$ will generically escape region $R_1$ at a time $T_1^{*}$. Taking  $T_1<T_1^{*}$, we have $y^* = \Psi_1(T_1, x^*) \in R_1$, and $y^*$ is above and to the right of $x^*$ (Figs. \ref{fig:phase1Lv}b and \ref{fig:phase1Lv}c).

\begin{figure}[ht]
	
	\centering
 
	\includegraphics[width=\linewidth]{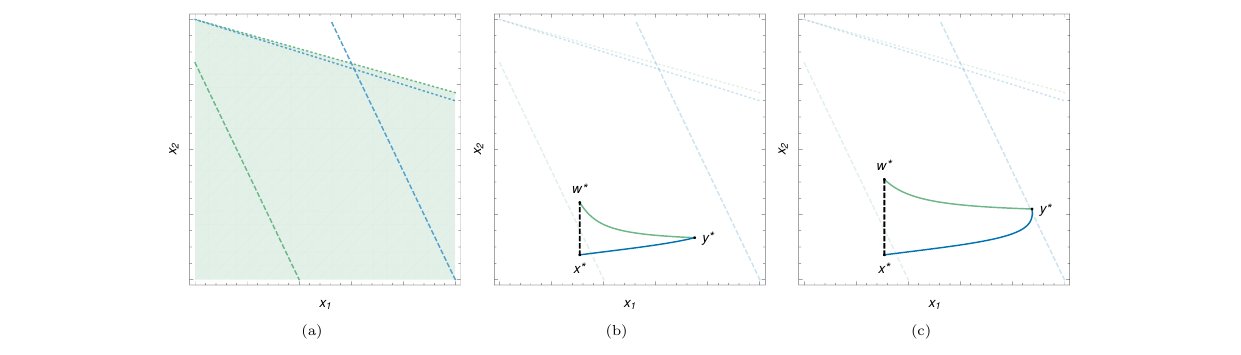}
	
	\caption{Regions $R_2$ (a) and $R^*$ limited by the curves connecting the points $x^*$, $y^*$ and $w^*$, considering different choices for $T_1$, the first phase duration (b, c). The orbits for the first and second phases are shown in blue and green, respectively.}
    \label{fig:phase2Lv}
\end{figure}

During the second phase, we administer drug $D_1$, which is effective against type 1, but not against type 2 cells. It is reasonable, then, to assume that the flow diagram of the second phase is of the type as in Fig. \ref{fig:possibleLV}d, i.e. we would observe the \textit{extinction of the sensitive cells} and survival only of the resistant cells if this treatment were used for infinite time. Geometrically, this implies that the $x_1$ nullcline must be below the $x_2$ nullcline in the first quadrant during the second phase, as in Fig. \ref{fig:possibleLV}d.  We define $R_2$ as the region below the $x_2$ nullcline during second phase of the treatment. In other words, $R_2$ is the region in which population $x_2$ increases during the administration of $D_1$, see Fig. \ref{fig:phase2Lv}a. We further assume that treatment with $D_1$ \textit{weakens sensitive cells} by decreasing their carrying capacity and their aggressiveness, but \textit{does not affect} the same quantities for the resistant ones. It follows that $R_1 \subset R_2$. We see that every tumor in $R_2$ would eventually become resistant to $D_1$ if we treat it only with this drug.

The above assumptions about the second phase of the treatment may be mathematically summarized as: 
\begin{subequations}
	\quad 
	
	$\bullet$ \textbf{Extinction of the sensitive cell population: }
	
	\begin{equation}
		k_{1, 2} < \frac{k_{2, 2}}{a_{21, 2}}\;\;\;\mathrm{and}\;\;\; \frac{k_{1, 2}}{a_{12, 2}} < k_{2, 2} \;.
		\label{ctr:lvPhase2Extinction}
	\end{equation}
	\quad 
	
	$\bullet$ \textbf{Weakening of sensitive cells: }
	
	\begin{equation}
		k_{1, 2} < k_{1, 1} \;\;\;\mathrm{and}\;\;\; a_{21, 2} \leq a_{21, 1} \;.
		\label{ctr:lvPhase2Decreases}
	\end{equation}
	\quad 
	
	$\bullet$ \textbf{Resistant cells are not affected: }
	
	\begin{equation}
		k_{2, 2} = k_{2, 1} \;\;\;\mathrm{and}\;\;\; a_{12, 2} = a_{12, 1} \;.
		\label{ctr:lvPhase2NotAffected}
	\end{equation}
	\label{ctr:lvPhase2}
\end{subequations}

If $y^* = \Psi_1(T_1,x^*)$ as before, the above assumptions imply that the future orbit of $y^*$, i.e. the curve $\Psi_2(t,y^*)$, $t>0$, must cross a vertical line through $x^*$ at some point $w^*$ at some time $T_2^*>0$. More yet, $w^*=\Psi_2(T_2^*,y^*)$ must be \textit{above} $x^*$. We define $R^*$ as the region limited by the curves $\gamma_1 = \Psi_1([0, T_1],x^*)$, $\gamma_2 = \Psi_2([0, T_2^*],y^*)$ and the line segment between the points $x^*$ and $w^*$. The above curves and regions are illustrated at Figs. \ref{fig:phase2Lv}b and \ref{fig:phase2Lv}c.

Finally, during the third phase we administer drug $D_2$, which is effective against both cell types. Naturally, such a treatment makes sense if tumors that are still developing do not become larger than they would without any treatment. In geometrical terms, this means that both nullclines of the third phase of the treatment are below the nullclines of the first phase of the treatment. In terms of parameters, see Fig. \ref{fig:possibleLV}, this imposes the following constraints:

\begin{subequations}
	\quad 
	
	$\bullet$ \textbf{$x_1$ nullcline constraints: }	
	\begin{equation}
		\begin{split}
			k_{1, 3} < \min\left\{k_{1, 1},\; \frac{k_{2, 1}}{a_{21, 1}}\right\}\;\;\;\mathrm{and}\;\;\;
			\frac{k_{1, 3}}{a_{12, 3}} < \min\left\{k_{2, 1},\; \frac{k_{1, 1}}{a_{12, 1}}\right\}\;.
		\end{split}
	\end{equation}
	\quad 
	
	$\bullet$ \textbf{$x_2$ nullcline constraints: }	
	\begin{equation}
		\begin{split}
			k_{2, 3} < \min\left\{k_{2, 1},\; \frac{k_{1, 1}}{a_{12, 1}}\right\}\;\;\;\mathrm{and}\;\;\;
			\frac{k_{2, 3}}{a_{21, 3}} < \min\left\{k_{1, 1},\; \frac{k_{2, 1}}{a_{21, 1}}\right\}\;.
		\end{split}
	\end{equation}
	\label{ctr:lvPhase3}
\end{subequations}

We define $R_3$ as the region where both cell types \textit{decrease} under the treatment with $D_2$, i.e., the region above both nullclines in phase 3, and also $R = R_3 \cap R_1$. For a tumor composition $x^* \in R$, populations for both cell types increase in treatment holidays and both decrease when the tumor is treated with drug $D_2$. 

\begin{figure}[ht]
	\centering
 
    \includegraphics[width=\linewidth]{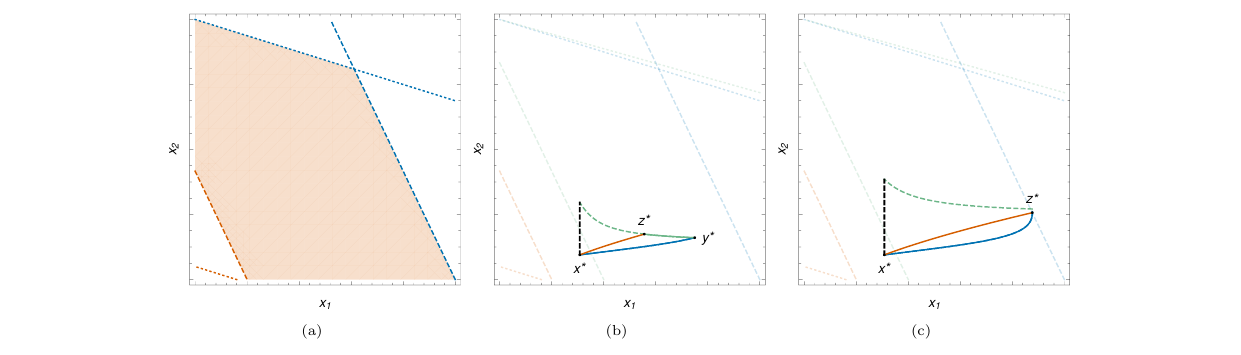}
	
	\caption{Region $R = R_1 \cap R_3$ (a) and cycles designed between the points $x^*$, $y^*$ and $z^*$ considering different times for the first phase (b, c). The orbits for the first, second and third phases are shown in blue, green and orange, respectively.}
    \label{fig:fullCycleLV}
\end{figure}

If $x^* \in R_3$ and we move \textit{backward} in time, then both cell types \textit{increase} indefinitely. This suggests a steepness condition necessary to guarantee existence of a cyclic routine: the past orbit of $x^*$ for the third phase is \textit{steeper} than the future orbit of $x^*$ for the first phase (i.e., the orange curve is, in the region close to $x^*$, above the blue curve as in Fig. \ref{fig:fullCycleLV}). 

In terms of derivatives, the steepness condition may be written as 
\begin{equation}
    \frac{\Psi_1^\prime(0, x^*)_2}{\Psi_1^\prime(0, x^*)_1} < \frac{-\Psi_3^\prime(0, x^*)_2}{-\Psi_3^\prime(0, x^*)_1}\;. \label{eq:inclinacaoLV}
\end{equation}

 Although providing valuable geometric information, the inequality above may be rewritten in order to convey more biological information. 
 In fact, $\Psi_1^\prime(0, x^*)_j$ is the growth rate of cell type $j$ in the absence of drugs, whereas $-\Psi_3^\prime(0, x^*)_j$ represents the death rate of cell type $j$ under treatment with $D_2$. Rewriting the above inequality as 
 \begin{equation}
		\frac{\Psi_1^\prime(0, x^*)_2}{-\Psi_3^\prime(0, x^*)_2} < \frac{\Psi_1^\prime(0, x^*)_1}{-\Psi_3^\prime(0, x^*)_1}\label{eq:inclinacaoLvBio}
	\end{equation} 
 reveals that it becomes a \textit{cost of resistance} condition stating that \textit{resistance of type 2 cells to drug $D_1$ comes at a cost either in cell proliferation in the absence of treatment, or in sensitivity to drug $D_2$}.

If the cost of resistance condition \eqref{eq:inclinacaoLvBio} holds, we choose $x^* \in R$, and parameters satisfy biological conditions \eqref{ctr:lvPhase2} and \eqref{ctr:lvPhase3}, then we can clearly design a cycle starting from $x^*$ with almost any duration $T_1$ for the first phase, subject only to $T_1<T_1^*$. Indeed, for small values of $t > 0$, \eqref{eq:inclinacaoLvBio} implies that the curve $\Psi_3(-t,x^*)$ enters region $R^*$. As both coordinates of $\Psi_3(-t,x^*)$ increase and tend to infinity as $t \to \infty$, the past orbit $\Psi_3(-t,x^*)$, $t>0$, will eventually leave $R^*$. Thus it must intersect the boundary of $R^*$ either at $\gamma_1$ (Fig. \ref{fig:fullCycleLV}c) or $\gamma_2$ (Figs. \ref{fig:fullCycleLV}b), forming a cyclic treatment. Durations $T_2$ and $T_3$ are such that $\Psi_3(-T_3,x^*)=\Psi_2(T_2,y^*)$.

If we want to avoid the atypical situation depicted in Fig. \ref{fig:fullCycleLV}c, in which $T_2=0$ and only drug $D_2$ is used, instead of taking $T_1<T_1^*$, it suffices to take $T_1 < \widetilde{T_1}$, where $\widetilde{T_1}$ is such that $\Psi_3(-t,x^*)$, $t>0$, is strictly above $\Psi_1(t,x^*)$ for $t \in [0,\widetilde{T_1}]$.

With the above considerations, we have derived the following theorem:
    \begin{thm}
		Assume that a tumor with two cell types under the three treatment phases as specified at the beginning of this subsection can be modeled by equations \eqref{eqs:systemLV} with the treatments imposing constraints \eqref{ctr:lvPhase2} and \eqref{ctr:lvPhase3} on the parameters. If $x^* \in R$ and the cost of resistance condition \eqref{eq:inclinacaoLvBio} holds, then there exists $\widetilde{T_1} > 0$ such that for every $T_1 \in (0, \widetilde{T_1})$ we can find unique $T_2, T_3 > 0$ such that $$\Psi_3(T_3, \Psi_2(T_2, \Psi_1(T_1, x^*))) = x^*.$$		\label{thm:cycleExistenceLV}
	\end{thm}

	\subsection{Adjusted Replicator Model}
    \label{subsec:repModel}

    Notice that if we consider the Lotka-Volterra model proposed on Subsection \ref{subsec:LVModel}, the tumor could be controlled using only drug $D_2$. On the other hand, as a continuous use of this drug would maintain lower populations of the type 1 and type 2 cells, it would possibly lead to the release of a third cell type, resistant to this treatment. 
    
    We take this into account by proposing a tumor model with three kinds of cancer cells: besides cells of types 1 and 2 as before, we assume here existence  of type 3 cells, resistant to $D_2$, but sensitive to $D_1$. We can still search for a routine composed by a treatment holiday (duration $T_1$), followed by administration of $D_1$ (duration $T_2$) and administration of $D_2$ (duration $T_3$).
    
 Since we are now on a three-dimensional context, building a cycle using our geometric approach in the Lotka-Volterra model is no longer feasible. We will proceed as in \cite{DuaMaNewton2021, ParkNewton2023}. 

Replicator dynamics \cite{taylorjonker, hofbauer1998evolutionary, Nowak2006} is the standard ODE system adopted in Evolutionary Game Theory. It is based on a \textit{game-theoretical} \textit{pay-off matrix} $B = (b_{ij})$. In the context of replicator dynamics, the variables $x_i$ now represent the \textit{fraction} of cells of type $i$ in the tumor, so we restrict our attention to the points in $S_2$, see Section \ref{sec:notations} for the notation. Again we find ourselves in a \textit{planar context}. The pay-off matrix is meant to represent the interactions among the different cell types, with entry $b_{ij}$ representing the benefit that a cell of type $i$ experiences when interacting with a cell of type $j$. Particularizing to the case of three cell types, the dynamics is defined by
   $$\begin{cases}
   	x_i^\prime = x_i\, (f_i(x) - \Phi(x)), \;\; i = 1, 2, 3\;,
    \end{cases}
    $$
where
\begin{equation}
    	f_i(x) = (Bx)_i \;,
    	\label{eq:fitnessDefgen}
    \end{equation} 
is the fitness of type $i$ and 
    $$\Phi(x) = \sum_{i=1}^3 x_i f_i(x)$$  
    is the mean population fitness.

Following \cite{DuaMaNewton2021} and \cite{ParkNewton2023}, we will use instead the \textit{adjusted} version of the replicator dynamics, i.e.
\begin{equation}
    	\begin{cases}
    		x_i^\prime = \displaystyle x_i\frac{f_i(x) - \Phi(x)}{\Phi(x)}, \;\; i = 1, 2, 3 \;.
    	\end{cases}
    	\label{eq:systemRep}
    \end{equation}  
It can be shown that if all elements of the pay-off matrix $B$ are positive, then the orbits of both dynamics are the same, although the speed in which orbits are traversed are different (see, for instance \cite{hofbauer1998evolutionary}). We will see in Subsection \ref{subsecMoran}, and it will be important for our purposes, that the \textit{adjusted} replicator dynamics has a sort of correspondence with the Moran process \cite{benaimweibull, traulsenetal2006}. The same does not hold for the \textit{standard} replicator dynamics. This is why we adopt from now on the adjusted replicator dynamics.

A mathematical property of the replicator dynamics is that orbits are not changed if constants are added to each column of the pay-off matrix \cite{hofbauer1998evolutionary, Nowak2006}. We mention this here in conjunction with the material in \ref{apx:changeModels}, where a pay-off matrix with some negative elements will appear. Positive constants will be summed to the columns of that matrix in order to make it positive.

Similar to \cite{DuaMaNewton2021, ParkNewton2023}, we write the pay-off matrix elements as
$$b_{ij}=1+ w_i a_{ij}\;,$$
where the $a_{ij}$ are all positive and the \textit{interaction coefficient} $w_i$ is a positive parameter representing the intensity of the interactions for cell type $i$. In terms of the new ``pay-off" matrix $A$ and of the interaction coefficients, the fitness of cell type $i$ is
    \begin{equation}
    	f_i(x) = 1 + w_i(Ax)_i \;.
    	\label{eq:fitnessDef}
    \end{equation} 
    
With assumptions $a_{ij}>0$ and $w_i \geq 0$, the expression above implies that the smallest fitness value is equal to 1 if $w_i=0$. As proposed in \cite{DuaMaNewton2021}, we consider a common interaction coefficient $w_0 > 0$ for all three cell types during the treatment holiday, when no drugs are used. The effects of the drugs on the cells will be introduced by decreasing the interaction coefficient of a cell type whenever this type is sensitive to the drug. Concretely, we assume that when treated with the largest concentration $C_i=1$ of a drug to which it is sensitive, a cell type will have its interaction coefficient equal 0.  Taking into account that cells of type $1$ are sensitive to both drugs $D_1$ and $D_2$, that cells of type $2$ are sensitive to $D_2$, but not to $D_1$, and cells of type 3 are sensitive to $D_1$, but not to $D_2$, we have
    \begin{subequations}
    	\begin{equation}
    		w_1 = w_0(1 - C_1 - C_2)
    	\end{equation}
    	\begin{equation}
    		w_2 = w_0(1 - C_2)
    	\end{equation}
    	\begin{equation}
    		w_3 = w_0(1 - C_1)\;.
    	\end{equation}
    	\label{eqs:drugEffectRep}
    \end{subequations}

To complete the specification of the dynamics, we only have to specify the values of $C_1$ and $C_2$ at each phase:
    \begin{itemize}
		\item \textbf{Phase 1} is a treatment holiday: $C_1=C_2=0$.
		\item During \textbf{phase 2} only drug $D_1$ is used: $C_1=1, C_2=0$.
		\item During \textbf{phase 3} only drug $D_2$ is used: $C_1=0, C_2=1$.
	\end{itemize}
 
 We start now, in analogy to what we did in the Lotka-Volterra model, to describe how to choose values for the durations $T_1$, $T_2$ and $T_3$ in order to construct a treatment cycle. We will also state the necessary mathematical and biological hypotheses needed to guarantee the existence of such cycles.
    
Similar to hypothesis \eqref{eq:inclinacaoLvBio} in the Lotka-Volterra model, it is natural to assume a cost of resistance and expect that sensitive cells, in the absence of any drugs, are fitter than any resistant cells. We suppose that this is the case, at least when we start the treatment. Without loss of generality, we may suppose also that cells of type $2$ are fitter than cells of type $3$. Our hypothesis on the pay-off matrix is thus that there exists a region $R_1 \subset S_2$ such that
    \begin{equation}
    	f_1(x) > f_2(x) \geq f_3(x)
    	\label{ctr:sensitiveFittest}
    \end{equation} 
for all $x \in R_1$ and that the initial state $x^*$ of the tumor is such that $x^* \in R_1$.
We need not impose any further hypothesis on the relative sizes of the pay-off matrix entries.
    
Let $\widetilde{T_1}$ be the time it takes for the future orbit $\Psi_1(t,x^*)$ to reach the border of $R_1$. If the orbit never reaches the border in finite time, we define $\widetilde{T_1}$ to be infinite.
    
\begin{figure}[ht]
    	\centering

        \includegraphics[width=\linewidth]{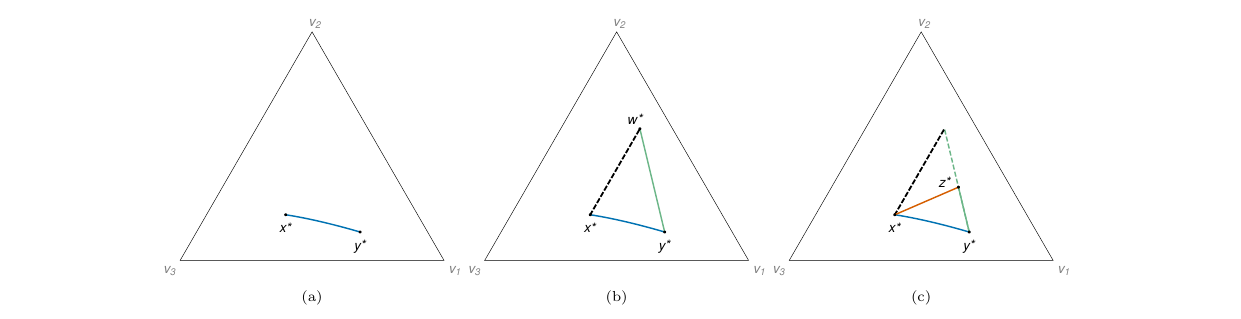}
    	
    	\caption{Cyclic routine designed between the points $x^*$, $y^*$ and $z^*$, with treatment orbits for the first (blue), second (green) and third (orange) phases. In panel (b) we can see the region $R^*$ limited by the curves connecting $x^*$, $y^*$ and $w^*$.}
        \label{fig:phasesRep}
    \end{figure}

We choose the duration $T_1>0$ of the first treatment phase to be arbitrary, but smaller than $\widetilde{T_1}$. Ordering in \eqref{ctr:sensitiveFittest} is thus preserved up to time $T_1$. This implies that during the first phase, the fraction $x_1$ of sensitive cells increases, while the fraction $x_3$ of the less fit cells decreases within the tumor (Fig. \ref{fig:phasesRep}a).  We define $y^*=\Psi_1(T_1,x^*)$, i.e. $y^*$ is the tumor composition at the end of the first phase.
    
During the second phase, the fraction $x_2$ increases whereas $x_1$ and $x_3$ decrease. The future orbit of any interior point of $S_2$ tends to the vertex $V_2$, because type 2 is the only cell type that can endure this treatment. This means that the future orbit of $y^*$ during the second phase must intercept the parallel to the $V_2V_3$ side of $S_2$ through $x^*$. We define $w^*$ as the intersection point of the future orbit of $y^*$ and the described parallel. A region $R^*$, similar to the one defined in the context of the Lotka-Volterra  model, can be defined, see Fig. \ref{fig:phasesRep}b, as the region bounded by the orbit section between $x^*$ and $y^*$, the orbit section between $y^*$ and $w^*$ and the line segment between $x^*$ and $w^*$. 
  
During the third phase, we administer the drug $D_2$ and, once again, it is natural to expect an increase and eventual fixation of type 3 cells, the only ones that are resistant to $D_2$. This implies that the \textit{past} orbit $\Psi_3(-t,x^*)$ tends to the face $V_1V_2$ side of the simplex as $t \rightarrow \infty$, with strictly increasing fractions $x_1$ and $x_2$.
    
Again, in analogy with the construction of a cyclic treatment in the Lotka-Volterra model, all we need in order to build a cycle here is to guarantee that the past orbit $\Psi_3(-t,x^*)$ enters region $R^*$. At difference with Lotka-Volterra, the slope condition \eqref{eq:inclinacaoLV} is automatically satisfied here. This is a direct consequence of the fact that $x^*$ satisfies \eqref{ctr:sensitiveFittest} (see \ref{apx:existenceRep}), without any further assumptions. This means that the curve $t \mapsto \Psi_3(x^*, -t)$ indeed enters the region $R^*$ and, since this curve goes towards the $V_1V_2$ side of $S_2$, it also leaves this region, forming a cycle (Fig. \ref{fig:phasesRep}c).
    
    With the above reasoning, we have proven the following    
    \begin{thm}
    	Consider a tumor with three cell types under the three treatment phases as specified at this subsection. Assume that the tumor can be modeled by equations \eqref{eq:systemRep}, that the pay-off matrix elements obey $a_{ij}>0$, and the interaction coefficients are subject to \eqref{eqs:drugEffectRep}. Finally, suppose that the initial tumor composition $x^*$ is in a region $R_1$ where \eqref{ctr:sensitiveFittest} holds. Then there exists $\widetilde{T_1} > 0$ (possibly $\infty$) such that for every $T_1 \in (0, \widetilde{T_1})$ we can find unique $T_2, T_3 > 0$ with $$\Psi_3(T_3, \Psi_2(T_2, \Psi_1(T_1, x^*))) = x^*.$$
    	
    	\label{thm:cycleExistenceRep}
    \end{thm}

    From a biological perspective, this theorem states that if we start a treatment in a state in which sensitive cells are the fittest ($x^*$ satisfies \eqref{ctr:sensitiveFittest}), and the cell types resistant to a drug become the fittest whenever this drug is administered, then we can design a cyclic treatment routine.

\subsection{Cycle stability}    
\label{subsec:Stability}
In case of application to real-life scenarios, the initial conditions for the cycles built in the preceding subsections are subject to small errors due e.g. to experimental errors in their determination. In the context of cyclic treatments, we want to understand how close to the planned we can expect the actual tumor composition to be throughout the treatment, under small perturbations in the cell populations. This is related to the topic of stability in dynamical systems.

Roughly speaking, a cyclic routine will be \textit{asymptotically stable} if its application drives perturbed initial conditions to the theoretical cycle, if the perturbation is sufficiently small. Instead, a cyclic routine will be \textit{unstable} if there exist solutions to the system starting as close as desired to the initial condition of the cycle which orbits flee away from the cycle. 

\begin{figure}[ht]
		\centering

        \includegraphics[width=\linewidth]{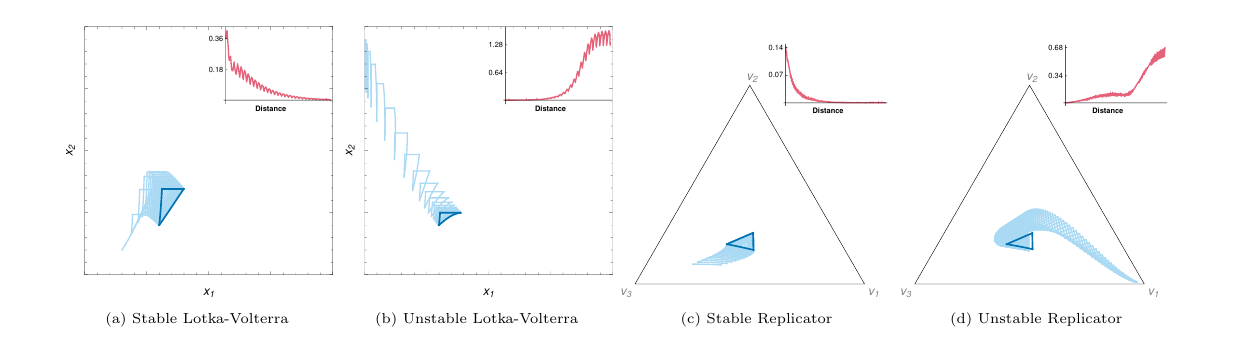}
		\caption{In dark blue, plots of orbits of treatment cycles created with the parameter values in Tables \ref{tab:lvParams} and \ref{tab:repParams}. In light blue, the orbits of perturbed initial conditions. In the top-right corner, the distance between the designed and the disturbed plots through time. The distance between the treatment initial condition and the perturbed one was taken as $\approx 0.0014$ for the unstable Lotka-Volterra routine and $\approx 0.024$ for the unstable replicator routine.} \label{fig:wholefigstableunstable}
	\end{figure}

In our numerical explorations with the cycles of the previous subsections for Lotka-Volterra and adjusted replicator dynamics, we found both stable and unstable cycles, for both dynamics. For examples of asymptotically stable cycles, see Figs. \ref{fig:wholefigstableunstable}a and \ref{fig:wholefigstableunstable}c, in which orbits for the cycles are shown, along with orbits of perturbed initial conditions that are attracted to them. For examples of unstable cycles, we refer to Figs. \ref{fig:wholefigstableunstable}b and \ref{fig:wholefigstableunstable}d. In these, we also show orbits of initial conditions very close to the cycle, but quickly going far from it. Parameter values producing the examples in Fig. \ref{fig:wholefigstableunstable} are given in Table \ref{tab:lvParams} for the Lotka-Volterra model and in Table \ref{tab:repParams} for the adjusted replicator model. 

 \begin{table}[htbp]
		\centering
		\begin{tabular}{|l||l|}
			\hline
			\multicolumn{2}{|c|}{\textbf{Lotka-Volterra Cyclic Routines}} \\
			\hline
			\hline
			\multicolumn{2}{|l|}{\textbf{Common Parameters:}}\\
			\multicolumn{2}{|l|}{$x^* = (0.6, 0.4)$} \\
			\multicolumn{2}{|l|}{$r_{1, 1} = 0.278;\ r_{1, 2} = 0.278 \times 10^{-3};\ r_{1, 3} = 0.278 \times 10^{-3}$} \\
			\multicolumn{2}{|l|}{$r_{2, 1} = 0.665;\ r_{2, 2} = 0.665 \times 10^{-3};\ r_{2, 3} = 0.665 \times 10^{-3}$} \\
			\multicolumn{2}{|l|}{$k_{1, 1} = 2;\ k_{1, 2} = 2 \times 10^{-3};\ k_{1, 3} = 10^{-2}$} \\
			\multicolumn{2}{|l|}{$k_{2, 1} = 2;\ k_{2, 2} = 2;\ k_{2, 3} = 2 \times 10^{-3}$} \\
			\hline
			\hline
			\textbf{Stable Parameters:} & \textbf{Unstable Parameters} \\
			$T_{1} = 2;\ T_{2} \approx 1.706;\ T_{3} \approx 1.535$ & $T_{1} = 3;\ T_{2} \approx 1.207;\ T_{3} \approx 0.419$ \\
			$a_{12, i} = 0.5;\ a_{21, i} = 0.9$ & $a_{12, i} = 1.5; a_{21, i} = 1.9$\\
			\hline
		\end{tabular}
		\caption{Specifications for Lotka-Volterra cyclic routines. Parameter values, except for the unstable aggressiveness, were based on the values given in \cite{Zhang2017}. The carrying capacities were scaled for a simpler transition to the adjusted replicator model, see \ref{apx:changeModels}, and the growth rates were scaled so that $T_1$, $T_2$ and $T_3$ have the same order of magnitude. }
			\label{tab:lvParams}
	\end{table}

Stability of solutions for Lotka-Volterra systems with periodic coefficients is a well-studied topic, see e.g. \cite{XiaHan} and references therein. In fact, Theorem 3.1 in that paper states conditions that guarantee that a competitive periodic Lotka-Volterra system has a unique periodic solution which is also \textit{globally} asymptotically stable. One of these conditions is that the related time-averaged Lotka-Volterra constant coefficients system has a globally asymptotic coexistence equilibrium, such as in Fig. \ref{fig:possibleLV}a. However Theorem 3.1 in \cite{XiaHan} does not strictly apply to the Lotka-Volterra cycles whose existence is described in the previous section, because the coefficients in our equations are discontinuous functions.

We did not find in the literature any results similar to Theorem 3.1 in \cite{XiaHan} for the adjusted replicator dynamics, but in our numerical explorations on adjusted replicator cycles built as in the preceding section, we also found both asymptotically stable and unstable cycles. 

Both for Lotka-Volterra and adjusted replicator dynamics, we conjecture that a cycle is asymptotically stable whenever the corresponding time-averaged system has an asymptotically stable coexistence equilibrium and unstable if the time-averaged system has an unstable coexistence equilibrium. 

\begin{table}[htbp]
		\centering
		\begin{tabular}{|l||l|}
			\hline
			\multicolumn{2}{|c|}{\textbf{Adjusted Replicator Cyclic Routines}} \\
			\hline
			\hline
			\multicolumn{2}{|l|}{\textbf{Common Parameters:}}\\
			\multicolumn{2}{|l|}{$x^* = (0.3, 0.2, 0.5)$} \\
			\multicolumn{2}{|l|}{$w_0 = 0.1$} \\
			\hline
			\hline
			\textbf{Stable Parameters:} & \textbf{Unstable Parameters} \\
			$T_{1} = 10;\ T_{2} \approx 5.759;\ T_{3} \approx 7.010$ & $T_{1} = 10;\ T_{2} \approx 5.287;\ T_{3} \approx 6.866$ \\
			&\\
			$A = \begin{bmatrix}
				1 & 1.8 & 1.8 \\
				0.8 & 0.8 & 1.1 \\
				0.8 & 1.1 & 0.8
			\end{bmatrix}$ & $A = \begin{bmatrix}
				1 & 1.8 & 1.8 \\
				0.5 & 1.1 & 1.3 \\
				0.5 & 0.8 & 1.2
			\end{bmatrix}$\\
			&\\
			\hline
		\end{tabular}
		\caption{Specifications for adjusted replicator cyclic routines. All parameter values, except for the stable matrix, were based on the values given in \cite{DuaMaNewton2021}. Due to the difference of our equation \eqref{eq:fitnessDef} with their analog equation, all entries in our unstable matrix are 1 less than theirs. Given the same initial condition and durations, the dynamics are the same.}
		\label{tab:repParams}
	\end{table}

The conjecture stated above is illustrated by the numerical examples appearing in Fig. \ref{fig:wholefigstableunstable}. It should be noted that both asymptotic stability and instability appear to be open properties, meaning that slight changes in the initial conditions, treatment durations, or model parameters do not alter these dynamical characteristics. From a practical standpoint, this suggests that the application of a routine, either stable or unstable, in real-life scenarios is likely to behave similarly to the theoretical predictions.

In summary, we may say that when using a deterministic model in designing an adaptive therapy cycle, it must be taken into account whether the cycle is stable or not. As we will see in the following, deterministic stability is also important if we choose a stochastic model.

	\section{Stochastic Models}
    \label{sec:stochasticModels}

\subsection{The Moran process as the stochastic counterpart of the adjusted replicator dynamics} \label{subsecMoran}

A well-suited stochastic model to use in conjunction with the adjusted replicator dynamics is the \textit{Moran process}. It was originally devised \cite{moran, Nowak2006} for populations with two types of individuals, but its extension to populations with three or more types of individuals was already studied \cite{Ferreira2020}. In this discrete-time stochastic process, we also consider a \textit{positive} pay-off matrix $A = (a_{ij})$ and selection coefficients $w_i$, but, instead of population fractions, we define $x_i$ as the absolute number of type $i$ individuals, as in the Lotka-Volterra model. At each time step one individual reproduces and one individual dies. The offspring of the reproducing individual has the same type of its parent and replaces the individual that died. The population size $N$ remains thus constant in time, although the numbers of individuals of each type may vary. Both reproducing and dying individuals are randomly drawn by independent lotteries. Whereas the death lottery is uniform, the reproduction lottery is such that the probability of an individual being drawn is proportional to its fitness. 
    
	The fitness of an individual of type $i$, to be used in the reproduction lottery, is defined \cite{Ferreira2020} as 
	$$f_i(x) = 1 + w_i \left(a_{ii}\frac{x_i-1}{N-1} + \sum_{j \neq i}a_{ij}\frac{x_j}{N-1}\right)\;,$$ 
	where $N = \sum x_i$ is the population size. Notice that \eqref{eq:fitnessDef} is recovered when $N \rightarrow \infty$.

	In order that fitter individuals have larger offspring, the probability that an individual of type $i$ is drawn for reproduction is $$\alpha_i(x) = \frac{x_i f_i(x)}{\sum x_j f_j(x)}\;.$$ On the other hand, every individual has the same probability of dying. Accordingly,$$\beta_i(x) = \frac{x_i}{N}$$ is the probability that a type $i$ individual dies.
	
	The relationship between an adjusted replicator dynamics model and a Moran process with the same pay-off matrix and interaction coefficients is known \cite{benaimweibull, traulsenetal2006}. Let $T$ be a finite time span. Consider the adjusted replicator dynamics solution at a certain initial condition for all $ t\in [0,T]$ and a realization of the Moran process with the same initial condition for $N T$ time steps. A general result, Lemma 1 in \cite{benaimweibull}, shows that the probability that the realization of the Moran process at step $k$ is far from the deterministic adjusted replicator solution in time $k/N$ tends to 0 as $N \rightarrow \infty$. In this sense, we may say that the Moran process ``converges" as $N \rightarrow \infty$ to the adjusted replicator dynamics, or that the Moran process is the stochastic counterpart of the adjusted replicator dynamics.

\begin{figure}[ht]
		\centering

        \includegraphics[width=\linewidth]{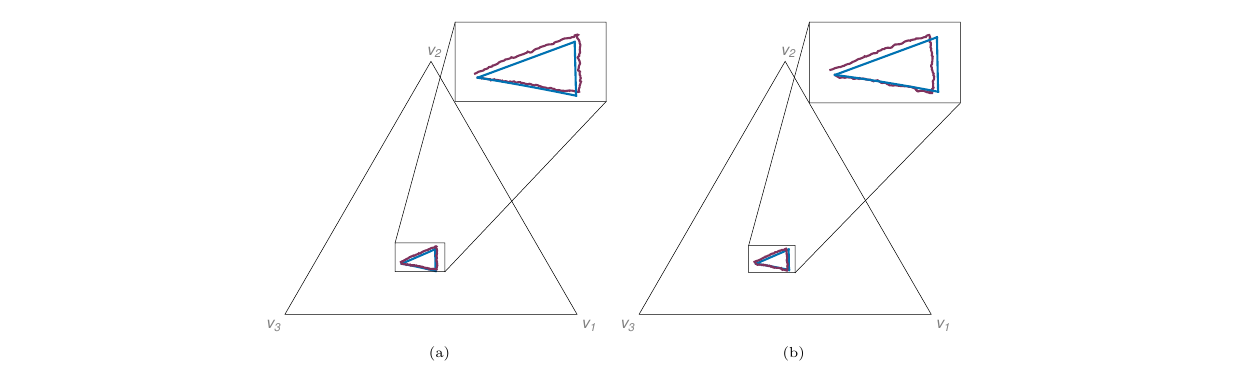}
  
		\caption{Realization of stochastic routines (purple) associated with the stable (a) and unstable (b) adjusted replicator routines (blue) designed with the parameters presented on Table \ref{tab:repParams} and $N=100,000$ cells. In the upper-right corner of each panel, we show magnified pictures of the most relevant part of the dynamics.}
		\label{fig:moranRep100k}
	\end{figure}

	A \textit{stochastic routine} may be defined for the Moran process by simply specifying its pay-off matrix and selection coefficients at each time step. Furthermore, given a deterministic routine designed with the adjusted replicator model, we can define an \textit{associated} stochastic routine by stating that the parameter values at time step $k$ correspond to the deterministic parameter values at time $k/N$. We may expect that, for large values of $N$, the associated stochastic and deterministic routines will remain \textit{close} for extended periods. But even if a deterministic routine is a cycle, i.e. its orbit returns \textit{exactly} to its initial point at the end of the cycle, we cannot expect that the corresponding stochastic routine does exactly the same. Fig. \ref{fig:moranRep100k} illustrates closeness between a deterministic routine and  its associated stochastic routine, both for the stable and the unstable cycles specified in Table \ref{tab:repParams}.
	
\begin{figure}[ht]
		\centering

        \includegraphics[width=\linewidth]{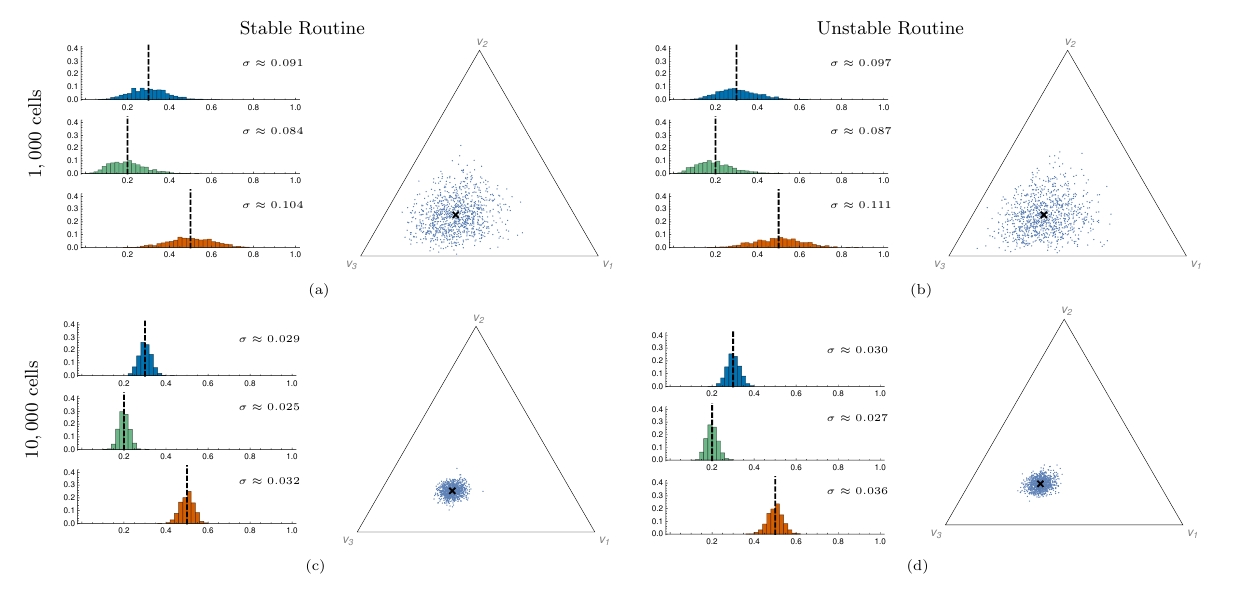}

		\caption{Spread of $1,000$ realizations of the stochastic routines associated with the deterministic cycles on Table \ref{tab:repParams} after one cycle and the distribution of each coordinate, $x_1$ in blue, $x_2$ in green and $x_3$ in orange. The initial state is marked with an $\times$. The standard deviation $\sigma$ for each distribution is shown with the corresponding histogram. (a) Stable stochastic routine with $N=1,000$ cells; (b) Unstable stochastic routine with $N=1,000$ cells; (c) Stable stochastic routine with $N=10,000$ cells; (d) Unstable stochastic routine with $N=10,000$ cells.}
		\label{fig:moranRep1Cycle}
	\end{figure}

The example realizations of the cycles in Fig. \ref{fig:moranRep100k} were run with a very large population size $N=100,000$ in order to showcase that the Moran process is an accurate stochastic counterpart for the adjusted replicator dynamics. In Fig. \ref{fig:moranRep1Cycle} we use instead smaller values of $N$ and we realize the same process $1,000$ times, as if we were using the same therapy in a cohort of $1,000$ individuals. As expected, the final state after one cycle \textit{spreads} around the initial state (marked with an \textbf{$\times$}) and the spread, measured by the standard deviation of the results, is larger in simulations with smaller population sizes.  At this point, the stability of the deterministic cycle does not seem to play a significant role on the robustness of the stochastic routine. Despite the spread, the initial condition is recovered in the average (Fig. \ref{fig:moranRep1Cycle}).

\begin{figure}[ht]
		\centering

        \includegraphics[width=\linewidth]{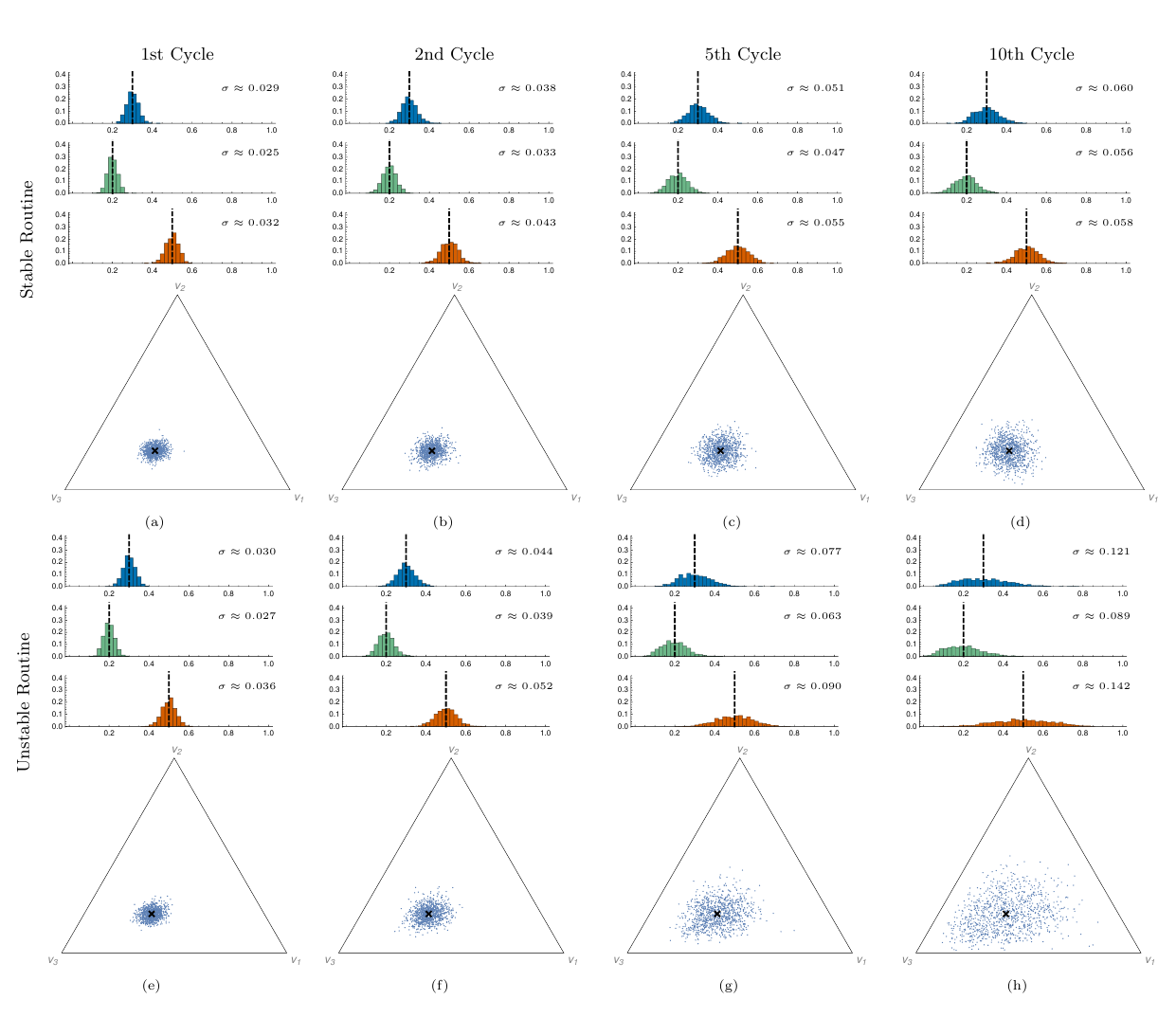}
		
		\caption{Spread of $1,000$ realizations of the stochastic routines associated with the deterministic cycles on Table \ref{tab:repParams} with $N = 10,000$ cells after successive cycles and the distribution of each coordinate, $x_1$ in blue, $x_2$ in green and $x_3$ in orange. The initial state is marked with an $\times$. The standard deviation $\sigma$ for each distribution is shown with the corresponding histogram. (a) Stable stochastic routine after $1$ cycle; (b) Stable stochastic routine after $2$ cycles; (c) Stable stochastic routine after $5$ cycles; (d) Stable stochastic routine after $10$ cycles; (e) Unstable stochastic routine after $1$ cycle; (f) Unstable stochastic routine after $2$ cycles; (g) Unstable stochastic routine after $5$ cycles; (h) Unstable stochastic routine after $10$ cycles;}
		\label{fig:moranRep10k}
	\end{figure}

A large difference between the unstable and the stable routines appears if we look at several repetitions of the same routine, as it is usually the case in cancer therapy. In the unstable case, the underlying effect of the deterministic dynamics tends to amplify the small distances due to the stochastic spread. On the contrary, in the stable case the deterministic dynamics tends to decrease the stochastic spread, which is inevitably recreated as time goes by. As a consequence, the stochastic spread is expected to increase with each repetition of the cycle in the unstable case. In the stable case, we expect the stochastic spread to preserve some stability over long periods of time. Results in Fig. \ref{fig:moranRep10k} confirm these expectations.

A Moran process stochastic routine is a Markov process with finitely many states and three absorbing states, namely the fixation of each cell type. Therefore, this process, even if it is the stochastic counterpart of an asymptotically stable cycle, must eventually be absorbed in one of these states \cite{Ferreira2020}, meaning it cannot stay close to the deterministic cycle indefinitely. Remember that the result of \cite{benaimweibull} we quoted here is valid for a finite time span $T$.

Indeed, as observed in \cite{DuaMaNewton2021}, the deviation from the deterministic cycle and absorption of the Markov process can occur over just a few cycles. We stress that the example cycle in \cite{DuaMaNewton2021} seems to be \textit{unstable}. The pay-off matrix used there is equivalent to the unstable matrix in Table \ref{tab:repParams}. 

On the contrary, in our simulations of Moran processes associated with asymptotically stable cycles with large values of $N$, we never saw any occurrence of absorption, indicating that the mean absorption time is probably much larger than the number of simulated time steps. We conjecture that the mean time to fixation in one of the absorbing states increases exponentially with $N$ in the asymptotically stable case. This expectation is based on analogy with results in \cite{AntalScheuring}, rigorously proved in \cite{teseRosangela}, on mean fixation times for the Moran process for two types of individuals, in the case that a stable coexistence equilibrium for the replicator equation exists.
	
  Dua, Ma and Newton \cite{DuaMaNewton2021} state, pessimistically, that a highly accurate tumor model is necessary to design an effective and repeatable adaptive treatment routine, because the stochastic errors will quickly degrade its effectiveness. We may be more optimistic, because we saw here that the stochastic fluctuations will not degrade too much a routine if it is based on an \textit{asymptotically stable} cycle.
        
    In summary, we have shown that the deviation from the planned cycle in stochastic routines is influenced by the stability or instability of the deterministic counterpart, and that this effect is more pronounced for smaller population sizes. We saw that stochastic fluctuations do not rapidly degrade a stochastic routine if it is based on an asymptotically stable cycle.

	\subsection{Stochastic Lotka-Volterra Model}
    \label{subsec:stochasticLV}

A well-known mathematical result, see Theorem 7.5.1 in \cite{hofbauer1998evolutionary}, states that a Lotka-Volterra system with $n-1$ cell types is equivalent to an adjusted replicator system with $n$ cell types. This equivalence means that each orbit of a Lotka-Volterra system in  $\mathbb{R}^{n-1}_+$ is mapped to exactly one orbit of an adjusted replicator in the $S_{n-1}$ simplex, and vice-versa. Thus, given a Lotka-Volterra system with $n-1$ cell types, we may write a $n\times n$ pay-off matrix that defines the equivalent dynamics. The construction of the replicator dynamics that represents a Lotka-Volterra system is described in \ref{apx:changeModels}. Notice that this construction involves defining an extra population compartment in the adjusted replicator system. We will refer to this extra compartment as the \textit{fictitious} cell type.

As we already know that the adjusted replicator dynamics has the Moran process as its stochastic counterpart, the equivalence result above allows us to relate a Moran process with a Lotka-Volterra system, thus creating a stochastic counterpart to the cyclic routines for Lotka-Volterra dynamics. 

Concretely, the path to be followed in order to produce a stochastic counterpart to a Lotka-Volterra system is the following: 
\begin{enumerate}
    \item Start with a Lotka-Volterra system, for example a cyclic treatment routine such as those studied at Subsection \ref{subsec:LVModel}.
    \item Obtain an adjusted replicator system equivalent to the Lotka-Volterra, using the procedure explained in \ref{apx:changeModels}.
    \item We already know from Subsection \ref{subsecMoran}
 that there is a Moran process that is a stochastic counterpart to the adjusted replicator system of the preceding step. Notice that the Moran process introduces a parameter $N$ to be the total population size.
    \item The stochastic counterpart to the Lotka-Volterra system is obtained, using again the equivalence between Lotka-Volterra and adjusted replicator, by mapping the path of the Moran process back to the space of the Lotka-Volterra system.
\end{enumerate}

Before translating a Lotka-Volterra system to an equivalent adjusted replicator system, a \textit{scaling} procedure may be advisable, in order that the associated Moran process path is not too close to absorption in one of the faces of the simplex. This scaling was already mentioned in the caption of Table \ref{tab:lvParams} and is explained in more depth at \ref{apx:changeModels}. 

A final word, before presenting results on stochastic routines associated with the deterministic cycles studied in Subsection \ref{subsec:LVModel}, is that the stochastic counterpart to a Lotka-Volterra system inherits a total population size $N$ from the subjacent Moran process. Of course this total population $N$ includes the fictitious cells needed to make the passage from Lotka-Volterra to adjusted replicator. It cannot, thus, be interpreted as a real population size. In particular, the total population size is \textit{not} constant in a Lotka-Volterra system. As we have seen in Subsection \ref{subsecMoran} that $N$ is also a measure of the size of the stochastic fluctuations in the Moran process, we propose that the population size $N$ of the stochastic counterpart to a certain Lotka-Volterra system is interpreted no more as a real population size, but as an \textit{effective} population size related to the size of the stochastic fluctuations in the stochastic counterpart. 

\begin{figure}[ht]
		\centering

        \includegraphics[width=\linewidth]{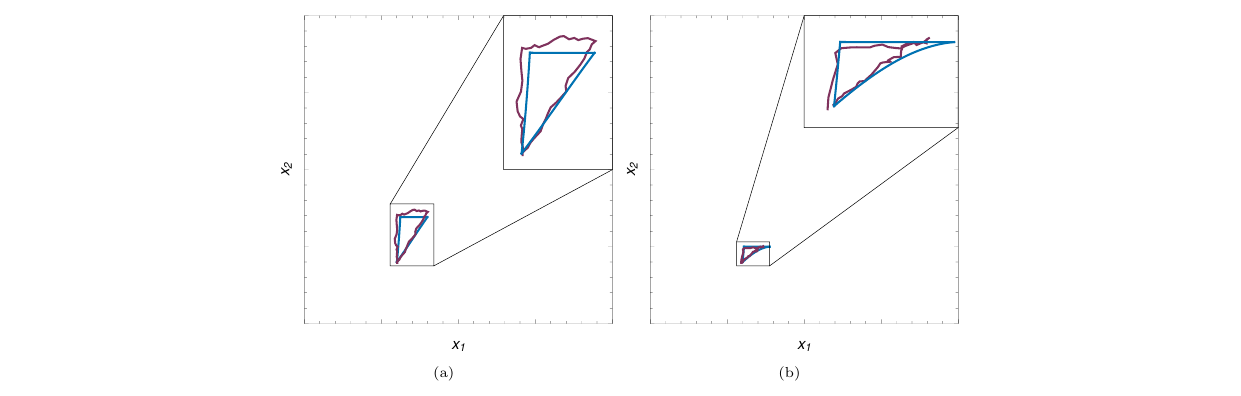}
  
		\caption{Realization of stochastic routines (purple) associated with the stable (a) and unstable (b) Lotka-Volterra routines (blue) designed with the parameters presented on Table \ref{tab:lvParams} with $100,000$ cells. In the upper-right corner of each panel, we show magnified pictures of the most relevant part of the dynamics.}
		\label{fig:moranLv100k}
	\end{figure}

In Fig. \ref{fig:moranLv100k}, we show realizations of stochastic routines associated with Lotka-Volterra systems. As expected, due to the large value of the effective population size $N$ used, we see that the stochastic paths remain close to the deterministic routines.
 
\begin{figure}[ht]
		\centering

        \includegraphics[width=\linewidth]{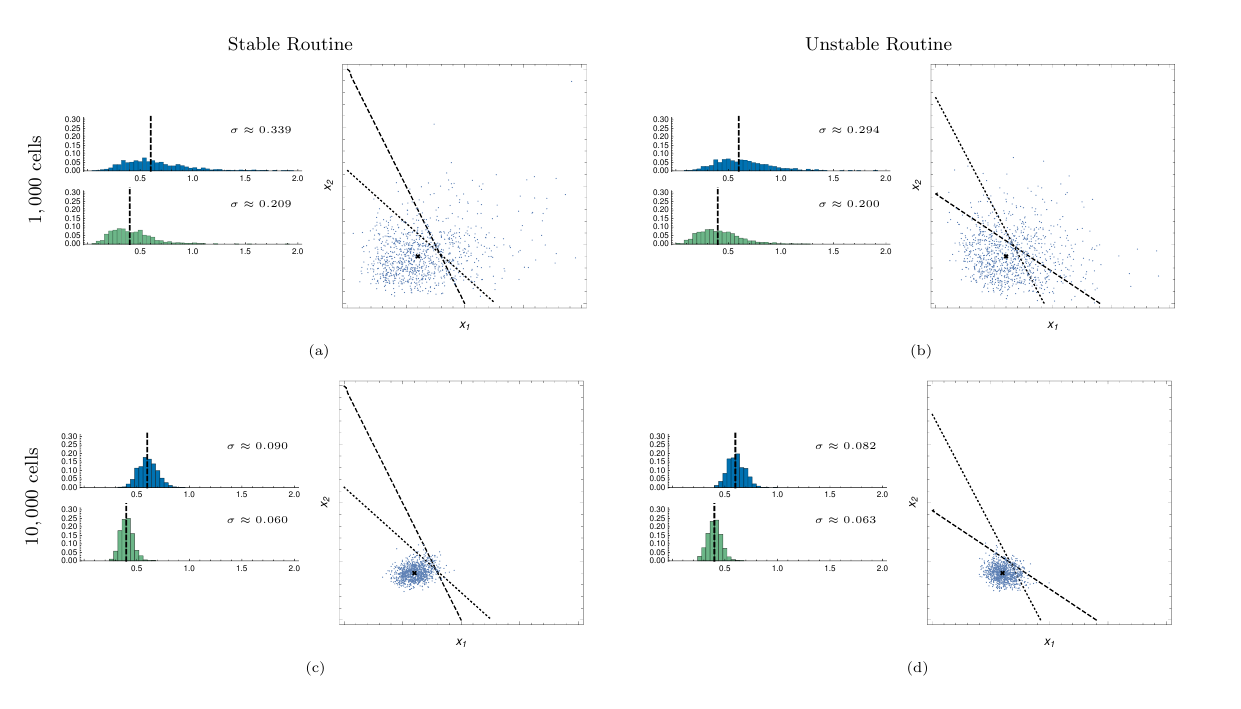}
		
		\caption{Spread of $1,000$ realizations of stochastic routines associated with the deterministic cycles on Table \ref{tab:lvParams} after one cycle and the distribution of each coordinate, $x_1$ in blue and $x_2$ in green. Also in the pictures, the nullclines relative to $x_1$ (dashed) and $x_2$ (dotted) for the treatment's time-averaged ODE system. The initial state is marked with an $\times$. The standard deviation $\sigma$ for each distribution is shown with the corresponding histogram. (a) Stable stochastic routine with $N=1,000$ cells; (b) Unstable stochastic routine with $N=1,000$ cells; (c) Stable stochastic routine with $N=10,000$ cells; (d) Unstable stochastic routine with $N=10,000$ cells;}
		\label{fig:moranLv1Cycle}
	\end{figure}

	The overall impact of the effective population size $N$ and the stability of the associated deterministic cycle are perceived in this model similarly to the adjusted replicator model. After one cycle (Fig. \ref{fig:moranLv1Cycle}), we notice that the final positions of many realizations of the same stochastic process spread over the plane and recover, on average, the initial condition. The standard deviation at this point is heavily impacted by the population size, but not by the cycle stability.

\begin{figure}[ht]
		\centering

        \includegraphics[width=\linewidth]{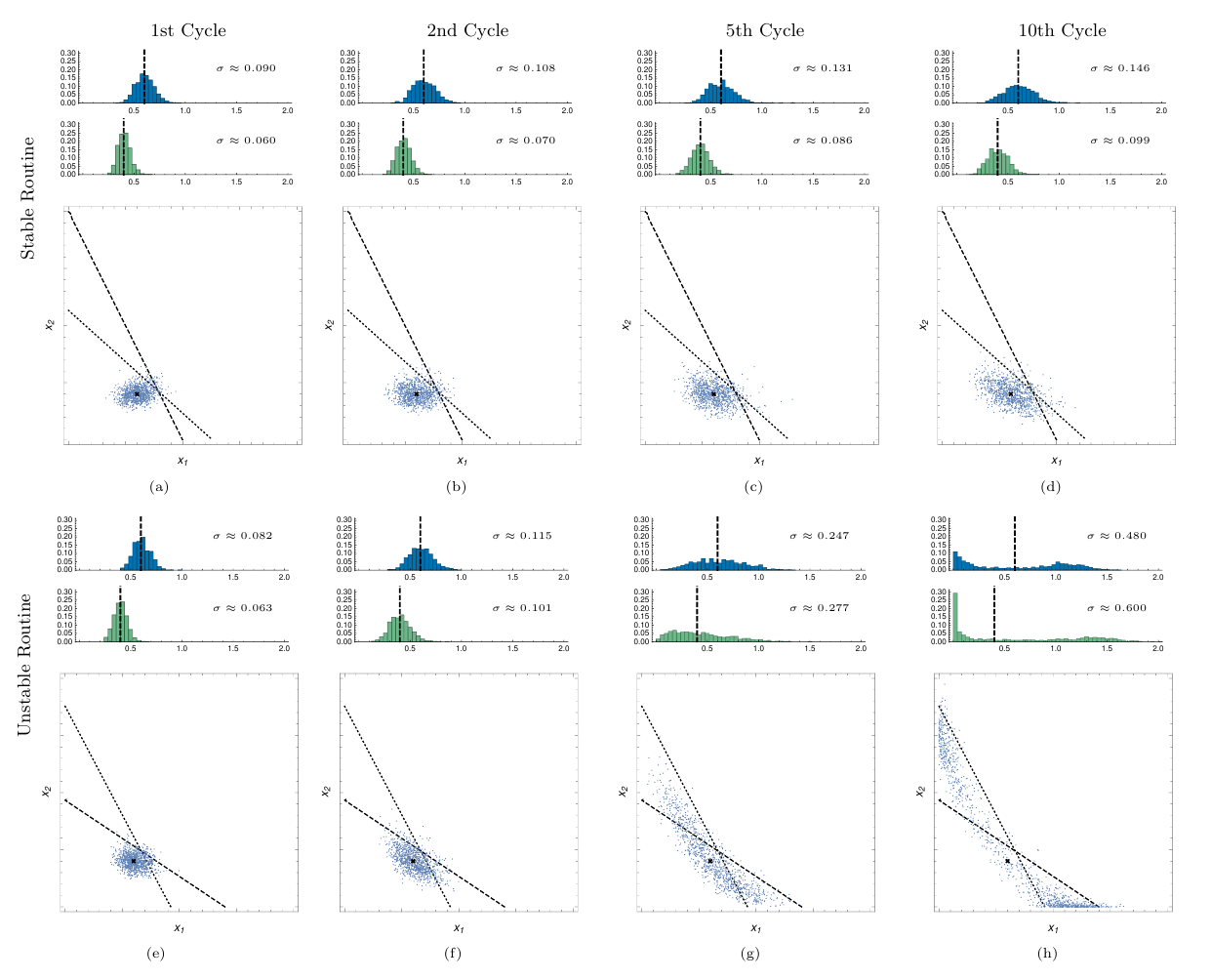}
		
		\caption{Spread of $1,000$ trials of stochastic routines associated with the deterministic cycles on Table \ref{tab:lvParams} with $N = 10,000$ cells after successive cycles and the distribution of each coordinate, $x_1$ in blue and $x_2$ in green. Also in the pictures, the nullclines relative to $x_1$ (dashed) and $x_2$ (dotted) for the treatment's time-averaged ODE system. The initial state is marked with an $\times$. The standard deviation $\sigma$ for each distribution is shown with the corresponding histogram. (a) Stable stochastic routine after $1$ cycle; (b) Stable stochastic routine after $2$ cycles; (c) Stable stochastic routine after $5$ cycles; (d) Stable stochastic routine after $10$ cycles; (e) Unstable stochastic routine after $1$ cycle; (f) Unstable stochastic routine after $2$ cycles; (g) Unstable stochastic routine after $5$ cycles; (h) Unstable stochastic routine after $10$ cycles;}
		\label{fig:moranLv10k}
	\end{figure}

As we repeat the routine over several iterations, the influence of the stability of the deterministic cycle becomes evident. Indeed, Fig. \ref{fig:moranLv10k} shows the final points of $1,000$ realizations of the routines associated with the Lotka-Volterra cycles specified in Table \ref{tab:lvParams} after 1, 2, 5 and 10 repetitions of the routines. Once again, the stochastic fluctuations are enhanced by unstable deterministic dynamics and suppressed in the stable, leading to a much slower spread.
 
As in the adjusted replicator model, both the effective population size and the stability or instability of the associated deterministic dynamics are crucial factors that impact the overall outcome of the stochastic Lotka-Volterra routine: after the first cycle, the spread of the final points is primarily driven by the population size (Fig. \ref{fig:moranLv1Cycle}) and, as the number of cycles increases, the effects of stability become more evident (Fig. \ref{fig:moranLv10k}).

	\section{Discussion}
    \label{sec:conclusion}
    
    The integration of ecological and evolutionary concepts into the study of cancer treatments introduces new and promising possibilities to the field. Adaptive therapy offers a novel approach to controlling tumors and delaying or even preventing the onset of drug resistance. This can be achieved over long periods using cyclic treatment routines. There exists already an extensive literature exploring these routines in different cancer contexts and with different models but, even though they are already being translated into clinical trials successfully, the mathematical foundation behind this process lacks the proper attention.

    We demonstrate in this paper, both for Lotka-Volterra and adjusted replicator models for tumors, that the design of treatment cyclic routines is indeed feasible under very reasonable biological conditions. Furthermore, we show that there exists an abundance of such cycles, which can be leveraged to identify treatment strategies that are better tailored to individual patient conditions.

    Along with the proof of existence of cycles, we also provided an algorithm to design cyclic routines given a starting point and the time to be spent in the first treatment phase. From a mathematical perspective, the generality of this algorithm and the geometrical arguments used in the demonstrations are not only applicable to the Lotka-Volterra and adjusted replicator models discussed here, but can also be extended to other models that exhibit planar dynamics. 

    The stability of deterministic routines is also taken into account in this paper. Understanding the mechanisms that drive the stability, and, consequentially, the predictability of a model, is valuable knowledge that greatly impacts our capacity of translating it to real-world cases. We state here a conjecture that, if proved, will predict the stability of a given cycle in terms of the dynamics of the time-averaged system. We conjecture that \textit{if the averaged system possesses an asymptotically stable coexistence equilibrium, the cycle itself will also be asymptotically stable, while if the system has an unstable coexistence equilibrium, the cycle will be unstable as well}. Given that the stability of the averaged system is an open property, this further implies that the treatment is likely to behave in clinical trials as predicted by the theoretical model.

 Our stability conjecture has an interesting biological interpretation: a stable treatment cycle is possible whenever the treatment creates an average environment that makes possible the stable coexistence of all cell types.
    
Finally, we showcase here, with both mathematical reasoning and numeric experiments, that the stability or instability of a deterministic routine heavily impacts the expected outcome of the associated stochastic routine when the treatment is applied for long enough. While studies showing that stochastic fluctuations may cause failure of a treatment routine are present in the literature, our work establishes that routines based on stable cycles may be useful for longer. This discovery may enhance our ability to manage and optimize treatment outcome.

A first limitation of our results is that, although the methods to construct cycles may be applicable to other models, non-planar systems remain particularly challenging, since ODE orbits do not limit regions in population spaces with three or more dimensions. A second limitation is that the application of these ideas to real cases depends on reliable determination of parameter values. With the proposed algorithm, we may even think of calculating cycles for single-individual treatments. But the correct determination of parameters for each individual is not only the outcome of an optimization procedure. Most probably, it will also rely on clinic experience.

    \section*{CRediT}

    \textbf{Yuri Garcia Vilela:} Methodology, Software, Formal analysis, Investigation, Data Curation, Writing - Original Draft, Writing - Review \& Editing, Visualization. \textbf{Artur César Fassoni:} Conceptualization, Writing - Review \& Editing. \textbf{Armando G. M. Neves:} Conceptualization, Methodology, Formal Analysis, Investigation, Writing - Review \& Editing.

    \section*{Funding}

    YGV is supported by the Coordenação de Aperfeiçoamento de Pessoal de Nível Superior - Brasil (CAPES) - Finance Code 001.

    ACF was supported by Alexander von Humboldt Foundation and Coordenação de Aperfeiçoamento de Pessoal de Nível Superior - Brasil (CAPES) - Finance Code 001, and partially supported by FAPEMIG RED-00133-21.

    AGMN is partially funded by FAPEMIG APQ-01784-22, Brazil.
 
	\appendix

	\section{Existence of Cyclic Routines in the Adjusted Replicator Model}
    \label{apx:existenceRep}
	
    In this section we present formal proofs for some statements that support Theorem \ref{thm:cycleExistenceRep}.
    
    As in Subsection \ref{subsec:repModel}, let us consider an initial condition $x^* \in \mathring{S_2}$ satisfying inequality \eqref{ctr:sensitiveFittest}. Directly from the inequality we have 
    $$\Phi(x^*) = x^*_1 f_1(x^*) + x^*_2 f_2(x^*) + x^*_3 f_3(x^*) < (x^*_1 + x^*_3 + x^*_3)f_1(x^*) = f_1(x^*)\;,$$ 
    therefore 
    $$\Psi_1^\prime(0,x^*)_1 = x_1^*\, \frac{f_1(x^*) - \Phi(x^*)}{\Phi(x^*)} > 0 \;,$$ 
    which implies that, indeed, the sensitive cell proportion is \textit{increasing} when we start the treatment. Analogously, we conclude that the $x_3$ proportion is \textit{decreasing} when we start the treatment and it follows that there exists $\widetilde{T_1}$ such that $y^* = \Psi_1( T_1^*,x^*)$ satisfies 
    $$y^*_1 > x^*_1, \ \ \ y^*_3 < x^*_3$$ for any $T_1^* \in (0, \widetilde{T_1})$.

  During the second phase of the treatment, drug concentrations are $C_1 = 1$ and $C_2 = 0$. Therefore, by \eqref{eqs:drugEffectRep}, we have $w_1 = w_3 = 0$ and, consequently, $f_1 \equiv f_3 \equiv 1$. On the other hand, as $x_2$ is resistant to the applied drug $D_1$, we also have $w_2 = w_0 > 0$ so 
    $$f_2(x) = 1 + w_0(Ax)_2 > 1 \;$$
    for all $x \in \mathring{S_2}$. Then, as shown for the first phase, the $x_2$ fraction increases while the $x_1$ and $x_3$ fractions decrease along the second phase. In particular, this implies that $\Psi_2(t,y^*)_2 \to 1$ and $\Psi_2(t,y^*)_1 \to 0$ when $t \to \infty$. It follows that the future orbit of $y^*$ under $\Psi_2$ will intercept the parallel to the $V_2V_3$ side through $x^*$ at a time $T_2^*$.  Letting $w^* = \Psi_2(T_2^*, y^*)$ be the intersection point, it satisfies $w_1^* = x_1^*$ and $w_3^* < y_3^* < x_3^*$. It also follows that $$w_2^* = 1 - w_1^* - w_3^* > 1 - x_1^* - x_3^* = x_2^*,$$ which justifies the construction of $R^*$ as in the Lotka-Volterra model (see Fig. \ref{fig:phasesRep}b).

   Finally, we demonstrate that condition \eqref{eq:inclinacaoLV} is a consequence of $x^*$ satisfying \eqref{ctr:sensitiveFittest} in this model. Indeed, notice that, as $f_1(x^*) > f_2(x^*)$ we have $$\frac{\Psi_1^\prime(0, x^*)_2}{\Psi_1^\prime(0, x^*)_1} = \frac{x^*_2(f_2(x^*) - \Phi(x^*))}{x^*_1(f_1(x^*) - \Phi(x^*))} < \frac{x^*_2}{x^*_1}\;.$$ 
    Meanwhile, on phase 3, we have $f_1 \equiv f_2 \equiv 1$, so 
    $$\frac{-\Psi_3^\prime(0, x^*)_2}{-\Psi_3^\prime(0, x^*)_1} = \frac{x^*_2(1 - \Phi(x^*))}{x^*_1(1 - \Phi(x^*))} = \frac{x_2^*}{x_1^*}\;.$$ 
    We conclude then, as stated, that 
    $$\frac{\Psi_1^\prime(0, x^*)_2}{\Psi_1^\prime(0, x^*)_1} < \frac{-\Psi_3^\prime(0, x^*)_2}{-\Psi_3^\prime(0, x^*)_1}\;.$$

	 \section{Equivalence of Lotka-Volterra and Adjusted Replicator Dynamics. Scaling Populations in Lotka-Volterra Systems} \label{apx:changeModels}
It can be shown, see Theorem 7.5.1 in \cite{hofbauer1998evolutionary}, that a general Lotka-Volterra system with $n-1$ species is equivalent to an adjusted replicator equation system with $n$ species. Equivalence means here that the orbits in $\mathbb{R}^{n-1}_+$ of the first can be mapped to orbits in $S_{n-1}$ of the second and vice-versa. In what follows we will describe the map taking one system to the other and the pay-off matrix that describes the adjusted replicator system in the special case of a competitive Lotka-Volterra system.

Let us consider a tumor composed of $n-1$ cell types with population given by $x = (x_1, \cdots, x_{n-1})$ and modeled with competitive Lotka-Volterra equations: \begin{equation}
	     \begin{cases}
	     \displaystyle x_i^\prime = r_i x_i \left(1 - \frac{\sum_{j=1}^{n-1} a_{ij}x_j}{k_i}\right), \;\;\; i = 1, \dots, n-1\;.
	 \end{cases}
     \label{eq:lvApx}
	 \end{equation}
If $a_{ii} = 1$, $i=1, \dots, n-1$, the $k_i$ are interpreted as carrying capacity of the $i$-th cell type, the $r_i$ are Malthusian growth rates, and $a_{ij}$, $i\neq j$ is the aggressiveness of type $j$ to type $i$. We also add to the population a ``fictitious cell type" with constant population conventionally chosen as $x_n \equiv 1$.

If $x \in \mathbb{R}^{n-1}_+$, we define $y \in S_{n-1}$ by 
\begin{equation}
   y_i= \frac{x_i}{\sum_{j=1}^n x_j}\;,
   \label{eq:LVtoRep}
\end{equation}
for $i = 1, \cdots, n$. Observe that $x_n=1$ implies that $y_n \neq 0$. Vice-versa, given $y \in S_{n-1}$, with $y_n \neq 0$, we define $x \in \mathbb{R}^{n-1}_+$ by
\begin{equation}
  x_i= \frac{y_i}{y_n}\;,
  \label{eq:ReptoLV}
\end{equation}
$i = 1, \cdots, n$.

The equivalence result \cite{hofbauer1998evolutionary} is that if $x$ is a solution to \eqref{eq:lvApx}, and $y$ satisfies an adjusted replicator system of equations
\begin{equation}
         \begin{cases} \displaystyle y_i^\prime = y_i \, \frac{f_i(y) - \Phi(y)}{\Phi(y)},\; i = 1, \dots, n\end{cases},
         \label{eq:systemLvRep}
     \end{equation}  
     with pay-off matrix $B = (b_{ij})$ given by 
     $$b_{ij} = \begin{cases}
        \phi_j -\displaystyle \frac{r_i a_{ij}}{k_i},\; i, j = 1, \dots, n-1 \\
         \phi_j + r_i,\; i=1, \dots, n-1, \;j = n\\
         \phi_j, \; i=n
    \end{cases}\;,$$
    then the orbits of one system are taken to the orbits of the other system by the maps \eqref{eq:LVtoRep} and \eqref{eq:ReptoLV}. For completeness, in \eqref{eq:systemLvRep} $f_i= (Bx)_i$ is the fitness of type $i$ and $\Phi = \sum_j x_j f_j$ is the population mean fitness. As the elements of the pay-off matrix must be positive to relate the adjusted replicator equations to a Moran process, $\phi_j$ are positive constants chosen so that all $b_{ij}$ are positive. Such additive constants are not present in the equivalence result as enunciated in \cite{hofbauer1998evolutionary}, but they do not alter the orbits \cite{Nowak2006} of the adjusted replicator dynamics, as already commented in Subsection \ref{subsec:repModel}. In the simulations we did, which results are shown in Figs. \ref{fig:moranLv100k} to \ref{fig:moranLv10k}, we chose 
    $$\phi_j=\begin{cases}
        1+\max_{i\in\{1,2\}} \frac{\displaystyle{r_i a_{ij}}}{\displaystyle{k_i}},\;\;\; j=1,2 \\
        1, \;\;\; j=3
        \end{cases}\;,$$
        guaranteeing that all entries in the pay-off matrix $B$ of the adjusted replicator associated to a Lotka-Volterra are at least equal to 1, thus positive.

    It is important to stress that the orbits of the two systems are taken one to the other, but speeds along the orbits are not in general equal, so the time it takes to traverse a segment of one orbit is not the same to traverse the corresponding segment of the other.

We remind the reader that the above mentioned equivalence involves one extra \textit{fictitious} cell type with population size $x_n$ is conventionally chosen as $x_n=1$, besides the real population sizes $x_1, \dots, x_{n-1}$. The variables in the associated adjusted replicator dynamics are the population \textit{fractions}
$$y_i = \frac{x_i}{\sum_i x_i}\;.$$ 
If the real population sizes are of order of thousands of cells, such as in the examples in \cite{West2019LotkaVolterra}, the $y_n$ fraction in the above equations is close to 0. On the other hand, if the real population sizes were measured in some other unit, such as millions of cells, then they may be numbers much smaller than 1 and, in such a case, $y_n$ is very close to 1, whereas the other fractions are tiny. In the first case, the adjusted replicator orbit equivalent to Lotka-Volterra would lie very close to the $S_{n-1}$ face where $y_n=0$. In the second case, it would lie very close to the $S_{n-1}$ vertex where $y_n=1$. None of these situations is desirable, because the Moran process associated with the adjusted replicator dynamics would be, in both cases, close to absorption.

It would be very awkward if the Lotka-Volterra equations were so much dependent on the unit in which one chooses to measure population sizes. In fact, they are not. The reader can readily see that we may \textit{scale} by a factor $\alpha>0$ all populations in a Lotka-Volterra system such as \eqref{eq:lvApx}, if we also scale the carrying capacities. More exactly, we define
$$X_i = \alpha \, x_i\;.$$
For example, if $x_i$ are population sizes measured in the unit of cells and $\alpha= 10^{-3}$, then the $x_i$ are population sizes measured in \textit{thousands} of cells. The scaled populations $X_i$ may be obtained by solving a Lotka-Volterra system with exactly the same parameters, except for the carrying capacities, which must also be scaled, i.e. we just take
$$K_{i}= \alpha \, k_{i} \;.$$

It is exactly this sort of scaling what we mention in the caption of Table \ref{tab:lvParams}. In that caption we mean that the carrying capacity values given in \cite{Zhang2017} for the untreated phase were all multiplied by $\alpha = 2 \times 10^{-4}$. This value was chosen so that the real population sizes shown e.g. in Figs. \ref{fig:wholefigstableunstable}a and \ref{fig:wholefigstableunstable}b are of order of magnitude close to 1, i.e. comparable to $x_3=1$. By doing so, the adjusted replicator orbits equivalent to the cycles in Figs.  \ref{fig:wholefigstableunstable}a and \ref{fig:wholefigstableunstable}b are conveniently not too close either to the $V_1V_2$ side, or to the $V_3$ vertex.


\begin{thebibliography}{10}
	
	\bibitem{Aktipis}
	Athena Aktipis, Virginia~S.Y. Kwan, Kathryn~A. Johnson, Steven~L. Neuberg, and
	Carlo~C. Maley.
	\newblock Overlooking evolution: A systematic analysis of cancer relapse and
	therapeutic resistance research.
	\newblock {\em PLoS ONE}, 6, 2011.
	
	\bibitem{AntalScheuring}
	Tibor Antal and István Scheuring.
	\newblock Fixation of strategies for an evolutionary game in finite
	populations.
	\newblock {\em Bulletin of Mathematical Biology}, 68, 2006.
	
	\bibitem{benaimweibull}
	Michel Benaïm and Jörgen~W. Weibull.
	\newblock Deterministic approximation of stochastic evolution in games.
	\newblock {\em Econometrica}, 71, 2003.
	
	\bibitem{Bukowski}
	Karol Bukowski, Mateusz Kciuk, and Renata Kontek.
	\newblock Mechanisms of multidrug resistance in cancer chemotherapy.
	\newblock {\em International Journal of Molecular Sciences}, 21, 2020.
	
	\bibitem{Butcher}
	J.~C. Butcher.
	\newblock {\em Numerical Methods for Ordinary Differential Equations}.
	\newblock John Wiley And Sons, Ltd, 2016.
	
	\bibitem{DuaMaNewton2021}
	Rajvir Dua, Yongqian Ma, and Paul~K. Newton.
	\newblock Are adaptive chemotherapy schedules robust? a three-strategy
	stochastic evolutionary game theory model.
	\newblock {\em Cancers}, 13, 2021.
	
	\bibitem{Ferreira2020}
	Eliza~M. Ferreira and Armando~G.M. Neves.
	\newblock Fixation probabilities for the moran process with three or more
	strategies: general and coupling results.
	\newblock {\em Journal of Mathematical Biology}, 81, 2020.
	
	\bibitem{Gatenby}
	Robert~A. Gatenby, Ariosto~S. Silva, Robert~J. Gillies, and B.~Roy Frieden.
	\newblock Adaptive therapy.
	\newblock {\em Cancer Research}, 69, 2009.
	
	\bibitem{Gedye}
	Craig Gedye and Vishal Navani.
	\newblock Find the path of least resistance: Adaptive therapy to delay
	treatment failure and improve outcomes.
	\newblock {\em Biochimica et Biophysica Acta - Reviews on Cancer}, 1877, 2022.
	
	\bibitem{hofbauer1998evolutionary}
	J.~Hofbauer and K.~Sigmund.
	\newblock {\em Evolutionary Games and Population Dynamics}.
	\newblock Cambridge University Press, 1998.
	
	\bibitem{moran}
	P.~A.P. Moran.
	\newblock Random processes in genetics.
	\newblock {\em Mathematical Proceedings of the Cambridge Philosophical
		Society}, 54, 1958.
	
	\bibitem{murray}
	J~D Murray.
	\newblock {\em Mathematical Biology : I . An Introduction , Third Edition},
	volume~1.
	\newblock Springer New York, NY, 2002.
	
	\bibitem{Nowak2006}
	Martin~A. Nowak.
	\newblock {\em Evolutionary dynamics: exploring the equations of life}.
	\newblock Harvard University Press, 2006.
	
	\bibitem{ParkNewton2023}
	J.~Park and P.~K. Newton.
	\newblock Stochastic competitive release and adaptive chemotherapy.
	\newblock {\em Physical Review E}, 108, 2023.
	
	\bibitem{teseRosangela}
	Rosângela~A. Pires.
	\newblock {\em Análise assintótica dos tempos médios de fixação no
		Processo de Moran}.
	\newblock PhD thesis, Universidade Federal de Minas Gerais, 2022.
	
	\bibitem{CancerStats}
	Hyuna Sung, Jacques Ferlay, Rebecca~L. Siegel, Mathieu Laversanne, Isabelle
	Soerjomataram, Ahmedin Jemal, and Freddie Bray.
	\newblock Global cancer statistics 2020: Globocan estimates of incidence and
	mortality worldwide for 36 cancers in 185 countries.
	\newblock {\em CA: A Cancer Journal for Clinicians}, 71, 2021.
	
	\bibitem{taylorjonker}
	Peter~D. Taylor and Leo~B. Jonker.
	\newblock Evolutionary stable strategies and game dynamics.
	\newblock {\em Mathematical Biosciences}, 40, 1978.
	
	\bibitem{traulsenetal2006}
	Arne Traulsen, Jens~Christian Claussen, and Christoph Hauert.
	\newblock Coevolutionary dynamics in large, but finite populations.
	\newblock {\em Physical Review E - Statistical, Nonlinear, and Soft Matter
		Physics}, 74, 2006.
	
	\bibitem{Wang2021}
	Jiali Wang, Yixuan Zhang, Xiaoquan Liu, and Haochen Liu.
	\newblock Optimizing adaptive therapy based on the reachability to tumor
	resistant subpopulation.
	\newblock {\em Cancers}, 13, 2021.
	
	\bibitem{Adler}
	Jeffrey West, Fred Adler, Jill Gallaher, Maximilian Strobl, Renee
	Brady-Nicholls, Joel Brown, Mark Roberson-Tessi, Eunjung Kim, Robert Noble,
	Yannick Viossat, David Basanta, and Alexander~R.A. Anderson.
	\newblock A survey of open questions in adaptive therapy: Bridging mathematics
	and clinical translation.
	\newblock {\em eLife}, 12, 2023.
	
	\bibitem{West2018Replicator}
	Jeffrey West, Yongqian Ma, and Paul~K. Newton.
	\newblock Capitalizing on competition: An evolutionary model of competitive
	release in metastatic castration resistant prostate cancer treatment.
	\newblock {\em Journal of Theoretical Biology}, 455, 2018.
	
	\bibitem{West2020}
	Jeffrey West, Li~You, Jingsong Zhang, Robert~A. Gatenby, Joel~S. Brown, Paul~K.
	Newton, and Alexander~R.A. Anderson.
	\newblock Towards multidrug adaptive therapy.
	\newblock {\em Cancer Research}, 80, 2020.
	
	\bibitem{West2019LotkaVolterra}
	Jeffrey~B. West, Mina~N. Dinh, Joel~S. Brown, Jingsong Zhang, Alexander~R.
	Anderson, and Robert~A. Gatenby.
	\newblock Multidrug cancer therapy in metastatic castrate-resistant prostate
	cancer: An evolution-based strategy.
	\newblock {\em Clinical Cancer Research}, 25, 2019.
	
	\bibitem{XiaHan}
	Yonghui Xia and Maoan Han.
	\newblock New conditions on the existence and stability of periodic solution in
	lotka-volterra's population system.
	\newblock {\em SIAM Journal on Applied Mathematics}, 69, 2009.
	
	\bibitem{Zhang2017}
	Jingsong Zhang, Jessica~J. Cunningham, Joel~S. Brown, and Robert~A. Gatenby.
	\newblock Integrating evolutionary dynamics into treatment of metastatic
	castrate-resistant prostate cancer.
	\newblock {\em Nature Communications}, 8, 2017.
	
	\bibitem{Zhang2023}
	Lei Zhang, Jianli Ma, Lei Liu, Guozheng Li, Hui Li, Yi~Hao, Xin Zhang, Xin Ma,
	Yihai Chen, Jiale Wu, Xinheng Wang, Shuai Yang, and Shouping Xu.
	\newblock Adaptive therapy: a tumor therapy strategy based on darwinian
	evolution theory.
	\newblock {\em Critical Reviews in Oncology/Hematology}, 192, 2023.
	
\end{thebibliography}

\end{document}